# A particle-based approach for the prediction of grain microstructures in solidification processes

Salem Mosbah[1*], Rodrigo Gómez Vázquez[2,3], Constantin Zenz[2], Damien Tourret[4], Andreas Otto[2]

[1] *SOLIDIFICATION SAS, Sophia Antipolis, France*

[2] *E311-02 Research Unit of Photonic Technologies, Institute of Production Engineering and Photonic Technologies, TU Wien, Getreidemarkt 9/BA09, Vienna, Austria*

[3] *LKR Light Metals Technologies, Austrian Institute of Technology, Giefinggasse 2, Vienna, Austria*

[4] *IMDEA Materials, 28906, Madrid, Spain*

[*] *Corresponding author email: smosbah@solidification.io*

**Abstract**

Grain microstructures are crucial to the mechanical properties, performance, and often lifetime of metallic components. Hence, the prediction of grain microstructures emerging from solidification processes at relevant macroscopic scale is essential to the design or optimization of new alloys and processing conditions. Yet, despite the broad range of multi-scale models proposed so far, all of them suffer from computational limitations, such that advances from computational and algorithm perspectives remain needed. Here, we present a novel approach for tracking crystallographic solidification grain envelopes capable of predicting competitive growth scenarios and columnar-to-equiaxed transitions for stationary grains. The model relies on classical assumptions and equations in use in several broadly used and thoroughly validated approaches (e.g. cellular automata). Yet, our approach defines the grain envelope using Lagrangian particles and tracks their evolution using an algorithm and an implementation relying on scalable libraries and using modern CPU/GPU architectures. The model is used to simulate several benchmarks of increasing complexity, and the results are compared to analytical, experimental, and numerical results from literature for the purpose of model validation. To highlight the applicability to real-world processes and the possibility of coupling the model with existing physics-based simulation tools, the model is also (one-way) coupled with a multiphysics laser-material-interaction model to simulate competitive grain growth during laser beam welding of steel.

## 1. Introduction

Microstructures in metallic alloys occupy a pivotal point within the process–structure–property–performance paradigm [1]-[4]. Hence, the prediction of grain structures resulting from solidification processes – often the starting point of microstructure emergence – is essential to leverage *Integrated Computational Materials Engineering* [1]-[3] approaches. In this context, accurate yet fast numerical models incorporating all underlying phenomena at microstructure scale – namely transport of heat, mass, and solute species, as well as solid-liquid interface stability and growth kinetics – are key to a computationally-guided design strategy for achieving application-specific properties and performance [4].

Solidification models cover a broad range of length and time scales [5][6]. At the lowest scales (~ nm), first principles may be used to calculate phase diagrams [7], and atomistic models allow studying the structure and properties of solid-liquid interfaces (e.g. anisotropic excess free energy and kinetic coefficient) [8], but their computational cost remains prohibitive for simulations at the scale of a representative polycrystalline structure. At a slightly higher scale (~ μm), the phase-field (PF) method is the most efficient and versatile method to simulate the evolution of complex interface patterns, such as dendrites, eutectics, and more [9]-[11]. However, the need for an accurate discretization of the interface morphology (e.g. the local curvature of every single dendrite tip) imposes stringent scale limitations, even when using advanced algorithms – including parallelization [12]-[15], adaptive remeshing [14][15], or spectral solvers [16]-[19]. On the other end of the scale range, macroscopic scale models applicable at the scale of entire ingots (~ m) are most often based on volume-averaged balances of heat, solute, and momentum, coupled with fields representing the volume fraction of one or several phases [20]-[23], but they do not capture details of the polycrystalline structure. Therefore, a variety of intermediate-scale models have been proposed, for instance based on dendritic needle network [24][25], grain envelope [26][27], Monte Carlo (Potts) [28], phase-field [29], or cellular automaton (CA) [30]-[37] approaches.

Among these models, the CA approach stands out as a mature method, incorporating all essential underlying physics (i.e. nucleation, growth kinetics, alloy-specific solidification path, microstructure length scales), validated against a range of experimental conditions [32]-[36]. In a CA model, within a spatial grid of cubic cells, nucleation is addressed statistically via explicit seeding of nuclei with random orientations, and their growth proceeds by the progressive transitions of cells over *liquid*, *mushy* (*i.e.* a mixture of solid and interdendritic liquid within the grain envelope), and *solid* states. The grain envelope, i.e. the fictitious surface joining actively growing dendrite arms, is represented as polyhedral building blocks – e.g. octahedrons for cubic (fcc, bcc) crystals, whose vertices correspond to ⟨100⟩ preferred growth directions. The coupling with the macroscopic transport of mass, species, and heat is typically achieved by coupling with a Finite Element (FE) solver. CA-based models are applicable in three dimensions at macroscopic scale [33]-[36]. However, the exponential storage and computation requirements with the number of cells require strategies for the dynamical allocation of *active* (below *liquidus* temperature) and *inactive* (fully solid or liquid) cells. Indeed, the inherent algorithm complexity of the CA model, where a cell state depends upon its neighboring cells, scales as $O(n^2)$, where *n* is the number of cells. Moreover, post processing cost, e.g. for recovering the predicted grain structure, also increases dramatically with the number of cells.

Here, we present an alternative approach that can be applied to predict the grain structure during solidification processes, e.g. casting, welding, or additive manufacturing (AM). The model tracks the evolution of the theoretical envelope for each grain without solving the inner evolution of the dendritic features. We introduce a new point-based grain envelope tracking scheme, implemented using libraries optimized for modern computing CPU/GPU architectures to enable large scale industrially relevant simulations. The model is presented in Section 2. Section 3 provides an extensive review of benchmark results to validate the proposed approach. Section 4 showcases a couple of original applications, including laser beam welding, where the model is coupled with an existing multiphysics laser processing solver. The results and their implications are directly discussed within each subsection of Sections 3 and 4. Finally, a brief summary and some perspectives are provided in Section 5.

## 2. Solidification Model

The main objective of the model is to predict polycrystalline grain microstructure formation during solidification of a metallic alloy, and the resulting heterogeneities, e.g. solute segregation at grain boundaries (GBs), and texture, which critically affect the potentially heterogeneous mechanical properties of cast parts [39][40]. To do so, we need to build a model that incorporates and couples the relevant underlying physics of (a) solid nucleation (Section 2.1.1), (b) crystal growth (Section 2.1.2), and (c) interactions between growing crystallites and the transport (e.g. flow) of heat and species (Section 2.2). The model itself relies on well-accepted assumptions and equations common to several "mesoscale" modeling approaches (e.g. [24]-[27],[30]-[37]). In its current version, the model remains limited to the nucleation and growth of stationary grains (i.e. excluding the potential floatation or sedimentation of grains). The main novelty is the introduction of a particle-based grain envelope tracking scheme, presented in Section 2.1.2, which allows an efficient data handling and the use of existing scalable libraries. The resulting scheme is generic and can be coupled with any model for the complementary thermo-fluid problem. In this section, we present the main underlying assumptions and equations of the model, which we later test against benchmark cases from the literature (Section 3) and later illustrate original applications to melting and welding (Section 4).

### 2.1. Particle-based tracking of solidification and melting

#### 2.1.1. Nucleation

Since nucleation is a rare event that occurs at the scale of small clusters of atoms, macroscopic approaches cannot be fully predictive, which is why stochastic phenomenological models are typically used in phase-field [41] or cellular automaton [30][31] models. As homogeneous nucleation is irrelevant in practical applications, where heterogeneous nucleation is the main mechanism at play [39][40], we only consider heterogeneous nucleation. We use a nucleation model similar to that in the seminal CA model of [30]. Therein, random distributions of locations and associated nucleation undercooling are initially generated, typically following normal distributions throughout the whole domain, including boundaries. While the statistical distribution of nucleation undercoolings and the nuclei density are treated as calibration parameters, this method allows a direct correlation to available experimental measurements of cast ingot grain densities [40].

#### 2.1.2. Growth

A meshless grain envelope tracking approach is used based on the hierarchical nature of dendritic structures. Each point within the grain envelope is ascribed a velocity corresponding to that of a dendrite tip in the crystal preferred growth directions. In a Cartesian coordinate system ($\vec{e}_x, \vec{e}_y, \vec{e}_z$), for a cubic crystal (e.g. face centered cubic, fcc, or body centered cubic, bcc) these correspond to ⟨100⟩ directions, i.e., [100], [010], [001], [$\bar{1}$00], [0$\bar{1}$0] and [00$\bar{1}$]. For the sake of simplicity, here we limit the presentation to such crystal systems, but the same approach could be extended to other crystal systems, e.g. hexagonal close packed (hcp).

An isolated stagnant equiaxed grain growing into a homogeneously undercooled and supersaturated melt grows isotropically in its six ⟨100⟩ directions. During its growth, the grain shape evolution is equivalent to an isotropic re-scaling of its envelope approximated by an encapsulating octahedron. The scaling factor can be computed via a kinetic law relating the dendrite tip velocity to the undercooling computed at the farthest dendritic tips (i.e. the octahedron vertices). In this configuration, only six points are required to define the grain envelope, as depicted in Figure 1(a).

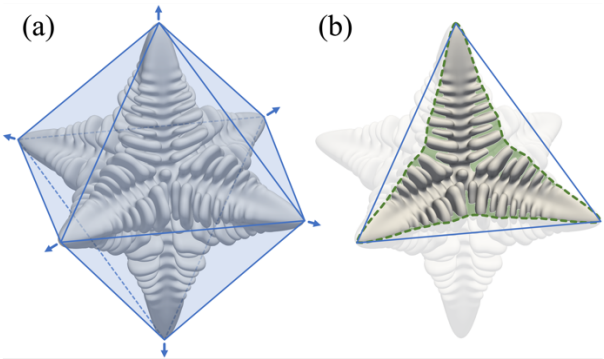

*Figure 1: Approximations of an equiaxed grain envelope: (a) octahedron represented with six tracking point growing along ⟨100⟩ directions (blue arrows); (b) comparison between schematic grain envelope (green shaded area and dashed line) joining the active secondary dendrite and its octahedral approximation (blue solid lines).*

In an inhomogeneous thermal and solute field, potentially in the presence of fluid flow, the grain envelope is still defined as the imaginary surface connecting the "active" dendrite tips. Active secondary (or *n*-ary) tips are usually defined as the tips of branches that are longer than any other secondary (*n*-ary) branch rooted closer to the tip of their common primary (*(n-1)*-ary) parent branch [26][27] as illustrated in Figure 1(b). Each envelope active dendrite tip, i.e. each tracking point, is expected to evolve following different growth kinetics based on its surrounding temperature and solute concentration. Under these conditions, the overall envelope can no longer be approximated by a simple polyhedron. Nevertheless, the actual grain envelope can be estimated by tracking the evolution of all the active dendrite tips. Therefore, an important consideration to accurately predict the shape of the grain envelope is the position of the outer primary, secondary and, if applicable, tertiary and higher-order dendritic tips. An optimal positioning of the tracking points is achieved when the distance between successive tracking points, denoted $\lambda_0$, represents a microstructural length scale of the dendritic pattern. A natural choice for $\lambda_0$ is the secondary dendrite arm spacing (SDAS), $\lambda_2$. Hence, in most cases we can use $\lambda_0 = \lambda_2$, but in practice the value of the numerical parameter $\lambda_0$ may be increased well above $\lambda_2$ in order to speed up the calculations, provided a convergence analysis on $\lambda_0$ (see, e.g., Section 3.3). Using existing scalable libraries (see Section 2.3), we implemented an algorithm to dynamically allocate the tracking points of the grain envelope.

The approach and implementation presented here are three-dimensional. For the sake of simplicity, we schematize a 2D growth case in Figure 2(a) and a generalized 3D case in Figure 2(b). Upon nucleation (Step 0), we can define an initial nucleus envelope from the nucleation center with six unique tracking points in 3D, or four tracking points in 2D (three of which are represented in Figure 2(a) illustrating only one half of a grain). These tracking points (smaller solid-filled points) are tracked up to a predefined target position (hatched points in Figure 2(a)), i.e. up to a distance $\lambda_0$. Once a tracking point reaches its target position, a new six/four points-based pattern is generated, and all points are initialized at the target position of the parent tracking point. The newly created points are each assigned a growth direction, initially forming an infinitesimal octahedral shape (or square in 2D). Only outer points, i.e. points growing in a direction that increases the overall grain volume, are considered. Hence, except for the initial nucleation event, the number of generated tracking points during growth is actually lower than six (lower than four in 2D). The position of each tracking point is then updated following the growth direction and the chosen kinetic law accounting from both local temperature and solute concentration fields. The process is repeated for each point resulting in a dynamically allocated, locally determined grain envelope.

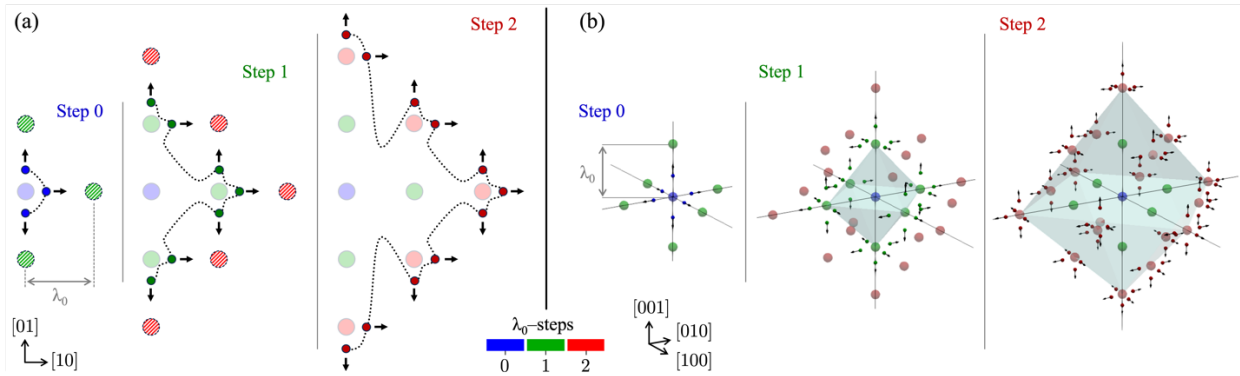

*Figure 2: Particle-based grain envelope tracking algorithm illustrated in 2D (a) and 3D (b) for an equiaxed grain growing into a homogeneously undercooled/supersaturated melt. At nucleation, active particles are initialized in each ⟨100⟩ direction. They grow following dendrite tip kinetics until they reach the next target position (hatched points in (a)), after which a new set of active tracking points is generated in outer ⟨100⟩ directions (arrows).*

Importantly, while Figure 2 illustrates the simplified case of symmetric (nearly-isotropic) equiaxed grains, since the growth of each tracking point depends on its local temperature and solute concentration fields, the approach naturally captures heterogeneous conditions, as shown later in Section 3.1. This contrasts with the simplest (early) CA algorithms where the growth of all vertices depends upon the driving force (undercooling or supersaturation) at the center of the cell [30][31][36], which was later addressed with slightly more advanced CA algorithms where the growth of each vertex can be calculated from its own surrounding driving force [32][37].

The resulting model, in a similar manner as a CA model, is thus capable of tracking the growth (and growth competition) of dendritic grains at macroscopic scale. In the case of polycrystalline growth competition, the advance of particles is naturally stopped when two particles from different grains meet, thus forming a GB. Since these envelope-based models do not explicitly account for individual dendritic branches, they cannot capture detailed mechanisms of dendrite impingement at converging GBs or sidebranching at diverging GBs, resulting in a simplified (nearly linear) morphology of resulting GBs [38]. However, in spite of their inability to accurately model the complex (rough) morphologies of GBs, once averaged over a sufficient population of grains (i.e. hundreds or more), such models nonetheless result in realistic statistical grain distributions in polycrystalline microstructures, as recently discussed based on a quantitative comparison between CA and PF simulations [38]. Morphological transitions, such as the columnar-to-equiaxed transition, hence emerge naturally from the applied nucleation and growth laws. Equiaxed grains appear whenever the nucleation criterion is fulfilled, their growth follows a similar kinetic law as the columnar grains that they compete with, and GBs are formed whenever two grain envelopes meet (regardless of whether they belong to columnar or equiaxed grains).

A key advantage of the particle-based approach over a traditional CA method is that it does not require to track the "capture" of neighboring cells (i.e. locating whether a grid point is located within a polyhedral shape) nor to calculate a new center for the new polyhedron associated with a newly captured cell [31][36]. Indeed, here we just track particles that essentially multiply when they reach a given distance $\lambda_0$.

The next missing component to the growth model is the correlation between the growth velocity, $v$, and local conditions, such as temperature, $T$, and/or local solute concentration, $w$. In practice, the particle-based approach can work with any expression of $v(T,w)$. Here, we simply use a classical KGT-type model [42][43], combining a local solute diffusion balance around the tip together with a tip selection criterion incorporating capillarity/curvature effect. For the solute transport, which may be assumed mostly diffusive in the immediate vicinity of the tip, we use the classical Ivantsov equation [44]. We consider a total undercooling $\Delta T = T_L - T$ below the Liquidus temperature $T_L$ of the alloy at its nominal concentration $w_0$. Kinetic undercooling is not considered here, since we focus on applications at moderate cooling rate (i.e. moderate growth velocity $\ll$ 1 m/s). As we aim for simplicity for a first demonstration of the particle-based approach, we chose a relatively simple set of equations for the growth kinetics model similar to the well-known KGT model for binary alloys [42][43] with two unknowns, namely the tip radius, $r$, and tip velocity, $v$. For its multicomponent extension, we use a simple additive contribution of the different solute species, as proposed and used in various models [35][49][50]. In practice, for the sake of computational efficiency, we use a power law $v(\Delta T)$ fitted to the KGT prediction of $v(\Delta T)$ for the given alloy, of the form

$$v = a\Delta T^b \tag{1}$$

or alternatively

$$v = a_1 \cdot \Delta T^{b_1} + a_2 \cdot \Delta T^{b_2} \tag{2}$$

with fitted coefficients $(a, b)$ or $(a_1, b_1, a_2, b_2)$.

### 2.1.3. Melting

Melting or re-melting can result from a local temperature increase beyond the Liquidus temperature of the alloy, due to melt flow, latent heat release during solidification, or an external energy source (e.g. a laser or electron beam). The present model handles melting by re-computing the new envelope with a "walk back" approach: *i.e.*, the solid/liquid interface is redefined by seeking among each of the previously defined tracking points those which result in a null undercooling. It is achieved via a search algorithm identifying the tracking points closest (within a predefined distance) to the iso-surface satisfying $\Delta T = 0$. The distance threshold is typically set as $2\lambda_0$. (Lower values lead to a faster search, but at the risk of a partially reconstructed discontinuous interface if chosen too low.) The model hence re-activates tracking points that are at the edge between positively and negatively undercooled liquid. Results presented here consider the Liquidus temperature of the alloy at its nominal concentration, but it would be relatively straightforward to extend it to a local composition-dependent Liquidus temperature. Melting is thus assumed here to result in a well-mixed interdendritic liquid, hence only requiring the value of the average composition, directly provided by the phase diagram. When an entire grain is melted, the original nucleation seed is kept in case of possible later re-solidification, however it is not allowed to interact with the grain structure until it reaches again the nucleation undercooling. This approach for melting and its implementation are tested and validated in Section 4.1.

## 2.2. Thermomechanical coupling

The particle-based grain growth model may be coupled with nearly any macroscopic thermo-mechanical model using a classical volume-averaging approach [20]-[23]. Here, we use a classical volume-averaged model and a coupling strategy similar to CA coupling schemes (see [33][34]). Below we briefly summarize the volume-averaged conservation equations (Section 2.2.1) as well as the coupling strategy (Section 2.2.2) used to obtain the results presented later in Section 3. Note that all mathematical symbols are summarized in Table S10 of the joint Supplementary Data.

### 2.2.1. Conservation equations

The conservation of mass is ensured with

$$\frac{\partial \rho}{\partial t} + \nabla \cdot (\rho \boldsymbol{U}) = 0, \tag{3}$$

where $\boldsymbol{U} = g^l \boldsymbol{U}^l + g^s \boldsymbol{U}^s$ is the volume-averaged flow velocity and $\rho = \rho(T)$ is the density of the mixture of solid and liquid. The liquid fraction, $g^l$, is defined as the sum of extra- and intra-dendritic liquid within a mushy zone of volume fraction, $g^m$, as

$$g^l = 1 - g^m g^{si} = 1 - g^s \tag{4}$$

Like in classical CA [30]-[37] models, the mushy zone is defined as a mixture of dendritic solid and interdendritic liquid. The mushy zone corresponds to the volume within the envelope of a grain, here delimited by the convex hull of the tracking particles for a grain. The internal solid volume fraction in the mushy zone, $g^{si}$, is expressed as a function of the local temperature, $T$, and concentration $w_i$ in solute species $i$ as

$$g^{si} = f(T, w_i), \tag{5}$$

which can be tabulated using different *solidification path* such as lever rule or Gulliver-Scheil model, either analytically or computationally using the CalPhaD method. Here, we used the lever rule, except for the simple scenarios in Sections 3.1, 3.3 and 4.1, where a linear variation of solid fraction with temperature was assumed and microsegregation was not accounted for.

The conservation of momentum follows the volume-averaged Navier-Stokes equations

$$\frac{\partial (\rho \boldsymbol{U})}{\partial t} + \nabla \cdot \left(\frac{\rho}{g^l} \boldsymbol{U}\boldsymbol{U}\right) = \nabla \cdot (\mu \nabla \boldsymbol{U}) - g^l \nabla p + g^l \rho \boldsymbol{g} - \frac{\mu}{K} g^l \boldsymbol{U} + \boldsymbol{S}_M . \tag{6}$$

The mixture density $\rho$ and viscosity $\mu$ are temperature-dependent, and the term $\boldsymbol{S}_M$ is the sum of source terms. The penultimate term represents the volumetric friction force where $K$ is the mushy zone permeability, here using the Carman–Kozeny relation:

$$K = \frac{\lambda_2^2 g^{l3}}{180(1-g^l)^2}, \tag{7}$$

derived based on the assumption of an isotropic porous mushy zone. Density gradients driven by species concentration are assumed to be dominant against temperature-induced density changes, such that the source term, $\boldsymbol{S}_M$, can be considered using the Boussinesq approximation:

$$\boldsymbol{S}_M = \rho_{T=T_L} \boldsymbol{g} \sum_{i=1}^{n} \{b_{w_i}(w_i^l - w_{i,0}^l)\}, \tag{8}$$

where $b_{w_i}$ are solute expansion coefficients for each species $i$.

The conservation of energy follows a classical diffusion-advection equation:

$$\frac{\partial \rho H}{\partial t} + \boldsymbol{U} \cdot \nabla \rho H - \nabla \cdot (\kappa \nabla T) = 0 \tag{9}$$

The mixture enthalpy, $H$, is defined as

$$H = H^s + g^l(H^l - H^s) \tag{10}$$

with

$$H^s = \int_{T_{ref}}^{T} c_p(T) dT \tag{11}$$

and

$$H^l = H^s + \Delta H_f, \tag{12}$$

where $c_p(T)$ is the heat capacity and $\Delta H_f$ is the enthalpy (latent heat) of fusion.

Species transport via convection and diffusion is modeled through

$$\frac{\partial w_i}{\partial t} + \boldsymbol{U} \cdot \nabla w_i^l - \nabla \cdot \left(D_i^l g^l \nabla w_i^l\right) - \nabla \cdot (D_i^s g^s \nabla w_i^s) = 0 \tag{13}$$

with the average composition, $w_i$, defined as

$$w_i = g^s w_i^s + g^l w_i^l \tag{14}$$

In most simulations presented here (i.e. all except the one in Section 4.2), the equations above are solved using a custom-built volume-averaged finite volume (FV) Navier-Stokes (NS) solver. Note that, in Section 3, we intentionally selected test cases that do not involve convection, so as to discard potential discrepancies expected from the resolution of the volume-averaged NS equation with different solvers [51], hence simply solving diffusion equations for enthalpy and solute concentration(s), when relevant. Furthermore, some test cases in Section 3 use a given temperature field as an imposed condition (namely Sections 3.2, 3.3, and 3.4). Material parameters (e.g. $D_i^s$, $D_i^l$, $c_p$, $\kappa$) may be phase- and temperature-dependent (in spite of their simplified notation here).

### 2.2.2. Coupling between macroscopic and microstructure model

Conservation Eqs (9) to (14) require an additional set of assumptions to compute the evolution of all fields. The heat conservation equation includes three unknowns, namely: the average temperature ($T$), the enthalpy ($H$), and the liquid fraction ($g^l$) fields. We use a classical enthalpy method [52] to solve the enthalpy as the primary unknown. We eliminate the temperature as an unknown using:

$$H^t - H^{t_0} = \frac{dH^*}{dT}(T^t - T^{t_0}) \tag{15}$$

with

$$\frac{dH^*}{dT} = c_p + \Delta H_f \frac{dg^{l^*}}{dT}, \tag{16}$$

where the superscripts $t_0$ and $*$ denote the previous time step and the previous iteration variable values, respectively. The term $dg^{l^*}/dT$ in the mushy zone is nonlinear and can be computed numerically or expressed analytically assuming a Gulliver-Scheil model or lever rule with a simplified phase diagram (e.g. with constant partition coefficient, $k$, and Liquidus slope, $m_L$).

A similar approach is used to isolate the average composition, $w_i$, for each species $i$ and solve the mass conservation equation as follows:

$$w_i^t - w_i^{t_0} \approx \frac{dw_i^*}{dw_i^l}\left[w_i^{l^t} - w_i^{l^{t_0}}\right] \tag{17}$$

with $dw_i^*/dw_i^l$ as the key term which is nonlinear in the mushy zone and can be derived from the considered solidification path (e.g. Gulliver-Scheil or lever rule). Equation (17), which is used to estimate $w_i^l$ in Eq. (13), relies on the fact that a local change of concentration in a finite volume comes from a change in liquid concentration, since the solid is here immobile.

In summary, the two-way coupling between the volume-averaged finite volume solver and the particle-based envelope tracking solver proceeds as follows. The evolution of the enthalpy (*H*), pressure (*p*), velocity (***U***), and average concentration (*w$_i$*) fields are solved via the FV solver from the system of Eqs (3), (6), (9), and (13). After the growth of the grain envelopes, i.e. the advance of the tracking particles following Eq. (1) or (2), the value of the mushy zone fraction (*g$^m$*) in each FV element is calculated from the location of the envelope. The mushy zone fraction is thus used in the FV solver to compute the solid fraction *g$^s$* and liquid equilibrium concentration $w_i^l$ via Eqs (4) and (14), accounting for a given solidification path (5), e.g. lever rule of Gulliver-Scheil model. These are subsequently combined to calculate the temperature field *T* via Eq. (10).

Up to this point, we have only addressed the growth of a single solid phase within an undercooled/supersaturated liquid. Yet, in most alloys, secondary reactions (e.g. eutectic, peritectic) are expected. In such cases, the considered solidification path, Eq. (5), could be substituted by a more advanced microsegregation model (e.g., [53][54]). Here, we use a simple approach whereby, once a given eutectic temperature *T$_{eut}$* is reached, all the enthalpy change is converted into a eutectic (solid) fraction, which does not directly affect the envelope growth model.

As already mentioned, the particle-based approach can be integrated with nearly any solver of arbitrary complexity. To illustrate this, in the final application (Section 4.2), it is (one-way) coupled to a multiphase model for the simulation of laser processing [55]. The solver accounts for fluid flow in both liquid and gas phases. It relies on a *mass-of-fluid* approach, which allows better mass conservation than the usual *volume-of-fluid* method in the case of compressible phases and phase change. Details on the governing equations, the numerical algorithm, and a series of results for validation purposes were recently presented elsewhere [55] and are therefore not repeated here.

### 2.3. Numerical resolution

Transport equations presented above are solved using the open-source C++-based finite volume (FV) solver OpenFOAM [56]. To couple the grain structure model with the continuum volume-averaged model, an accurate estimation of $g^l$, $dH/dT$, and $w^l$ is essential. These quantities are calculated based on the value of the mushy zone fraction, which is estimated at the particle positions from the cell centers of the FV mesh. While the FV solver calculates quantities (e.g. fields) at the center of cells, the local undercooling used to calculate the growth velocity for each particle point uses the local value of temperature and solute concentration(s) interpolated between cell centers (here using the OpenFOAM built-in interpolation scheme *cellPointFace*), which ensure continuity of undercooling across cell boundaries. We show later (Section 3) that this approach is adequate and accurate enough for a correct coupling while minimizing expensive interpolation operations.

For the particle-based grain growth model, we designed an algorithm that takes advantage of modern computer architectures and the use of scalable libraries. Namely, we use OpenVDB [57] space-partitioning acceleration structure, where active points are partitioned into voxels to accelerate search steps for particles within a given spatial range or for nearest neighbors. The resulting dynamic allocation strategy reduces the dendritic structure evolution problem to the computation of the evolution of a point cloud, where each point only carries essential information about grain properties (namely orientation and nucleation origin). Grain boundaries are naturally computed from serial, multi-threaded, and/or GPU-accelerated collision detection. Information about the grain inner volume is only necessary in case of re-melting and is stored using OpenVDB volumetric, dynamic grid that shares several characteristics with B-trees [57]. The result is a compact encoded data and grid topology representation of the solidification grain structure that requires minor overhead of CPU/GPU clock cycle and dynamic/persistent storage.

In essence, the model predicts the grain envelopes as a cloud of points. In order to render the grain structure, a Voronoi kernel-based interpolation can be used to recover the closest tracking point grain orientation at any given point. The Voronoi interpolation allows recovering spatial data within the empty shell-like point cloud for each grain, therefore allowing efficient data storage compared to CA-based models.

In summary, the underlying assumptions and equations of the model are somewhat standard, e.g. providing a grain growth description comparable to that in the popular CA method [30]-[36]. However, the particle-based description of the grain envelope and the choice of scalable tools and libraries results in a lighter data structure. The computational efficiency and a detailed scalability analysis will be discussed in a separate article (see preliminary results in [58]). First, in the present paper, we focus on the verification and validation of our model and numerical implementation against relevant existing benchmark cases from the literature (Section 3) and then illustrate the potential of the fully coupled model (Section 4).

## 3. Verification & validation against benchmark cases

In this section, we simulate different benchmark cases from the literature, quantitatively comparing the results to experimental data, analytical solutions, or numerical results, in order to verify and validate the proposed model and its numerical implementation. The cases are selected to test individually the capabilities of the model to simulate different key physical phenomena, with an increasing level of coupling and complexity, up to real-word applicable scenarios. For the sake of reproducibility, while maintaining the relative brevity and hence clarity of the main text, all input parameters such as material properties, initial and boundary conditions, and numerical discretization, are gathered in tables provided as Supplementary Material.

### *3.1. Single grain envelope in a temperature gradient*

First, we verify the accurate prediction of three-dimensional grain envelopes in a heterogeneous thermal field, by simulating the growth of a single grain in the presence of a temperature gradient, where an analytical solution exists [30]. Here, the simulation is only testing the particle growth algorithm, as it involves no solute field, and the temperature field is imposed. The grain orientation is defined by the Euler angles (20°, 20°, 20°) and the growth kinetics model follows Eq. (1) with $a = 10^{-4}$ m/(s K$^2$) and $b = 2$. The initial undercooling is set to $\Delta T = 2$ K, the cooling rate is $\dot{T} = -0.1$ K/s, with a thermal gradient of $G = 250$ K/m in the $\vec{e}_z$-direction. The microstructure length scale is set to $\lambda_0 = 200$ μm and the time step is $\Delta t = 0.01$ s. In Figure 3, we compare the model prediction at a time $t = 5$ s with the analytical solution provided in [30], showing a perfect match of the two solutions. Although the test case is very simple, it provides a valuable benchmark to verify the particle-based approach and its implementation in terms of accounting for the combined effect of the surrounding thermal field and crystal orientation.

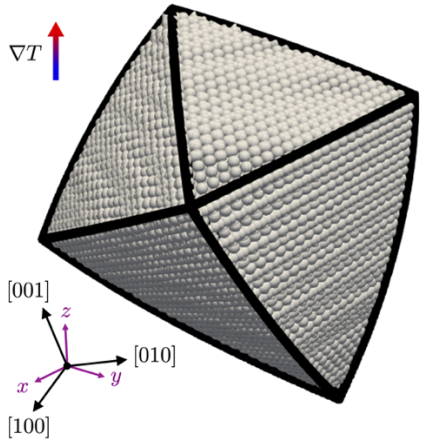

*Figure 3: Particle-based envelope tracking for a tilted grain within a temperature gradient. Comparison of the model prediction (spheres) and the analytical solution (black outline) provided by Gandin and Rappaz [30] of the grain envelope at t = 5 s after nucleation with an initial undercooling ΔT = 2K, a grain misorientation (ψ, θ, φ) = (20˚, 20˚, 20˚), a cooling rate of 0.1 K/s, and a thermal gradient G = 250 K/m in the $\vec{e}_z$-direction.*

### 3.2. Unidirectional solidification of an Al-Si alloy

Next, we test the model in the case of a one-dimensional (1D) unidirectional solidification of an Al-7wt.%-Si alloy and compare our results to that of Carozzani et al. [33]. In this case, the calculation involves only the diffusive thermal field coupled with the particle-based solidification model. One domain boundary obeys a heat flux boundary condition defined by a heat transfer coefficient and an outside temperature, while the other boundary is adiabatic. As our model is inherently three-dimensional, we simulate a 5 mm × 5 mm × 100 mm domain with adiabatic boundary conditions on the lateral (short dimension) directions. The material parameters, initial and boundary conditions, and discretization used in this simulation are provided in Table S1 of the joint supplementary document.

In Figure 4, we plot the time evolution of temperature and solid fraction at several probe locations, compared to the CA-FE model results presented in [33]. For consistency, we use as reference the non-iterative CA-FE algorithm described in [33]. Our results show an excellent agreement with these reference results. Discrepancies are of the same order as those observed between CA-FE and front tracking models (see Fig. 5 in [33]). As discussed in [33], the small oscillations and steps observed in the vicinity of Liquidus and eutectic temperatures (also observed in CA-FE results) are due to a change of slope in the enthalpy-temperature conversion, and they can be reduced with a finer spatiotemporal numerical discretization [33]. As also discussed in [33], while an iterative coupling scheme would allow a more accurate matching to front tracking models, the faster non-iterative algorithm leads to acceptable deviations (< 0.5˚C in this case), which are reduced even further in 3D compared to 1D and 2D simulations. These results show that the present model leads to predictions that are very close to those from a CA-based approach (even up to similar numerical oscillations).

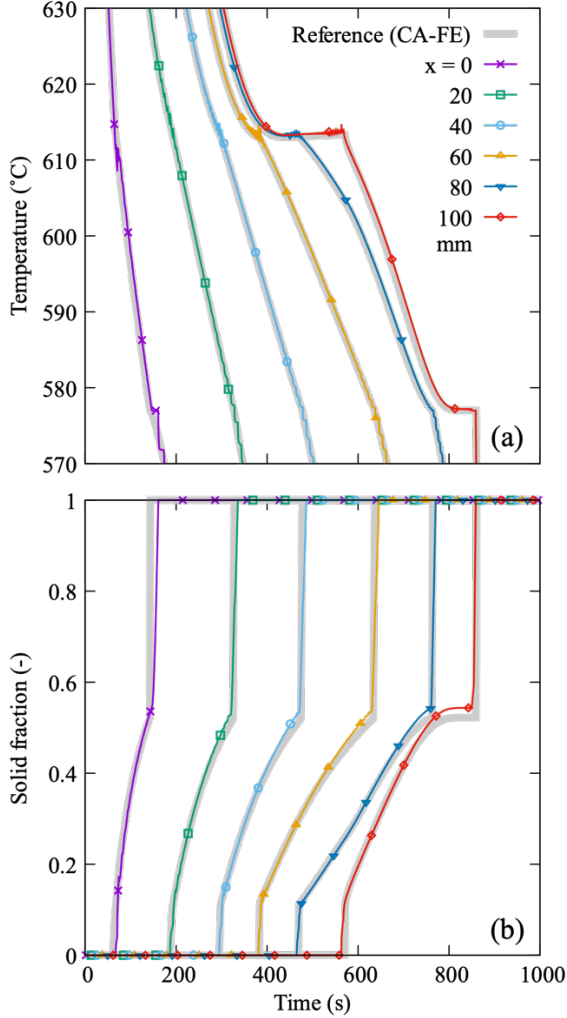

*Figure 4: One-dimensional simulation of Al-7wt.%Si solidification: (a) Temperature and (b) solid fraction at x = 0, 20, 40, 60, 80, and 100 mm, computed with the present model (foreground thin color lines with symbols) compared to the CA-FE results from [33] (background thick gray lines).*

### *3.3. Directional solidification of a quasi-2D thin sample*

Next, we test the model for a simulation of grain growth competition in a simple, two-dimensional case. The selection of grain boundary (GB) orientation during dendritic grain growth competition, even in 2D within a simple one-dimensional temperature gradient, is stochastic (due to the origin of sidebranches in the selective amplification of noise) [39][40][47]. However, models based on a CA or the model presented here are inherently deterministic. Here, we focus on one specific case in which the CA was previously shown to closely approach experimental measurements as the grid spacing is refined [30]. Namely, this corresponds to the pioneering experiment on a transparent succinonitrile-acetone alloy by Esaka et al. [61].

Here, the problem consists of a diffusive thermal field coupled with the particle-based solidification solver (no solute field). As the experiment contains three grains, we initialize our simulation by seeding 15 nuclei at the bottom of the domain at locations $(x,y) = (w/2+i\times 360, 0.15)$ μm with $w$ the width of the domain in the $y$-direction and $-7 \leq i \leq +7$. The three leftmost nuclei ($i \leq -5$) are initialized with a tilt angle $\psi = 4°$, the three rightmost nuclei ($i \geq 5$) with $\psi = 11°$, and the remaining central nuclei ($|i| \leq 4$) with $\psi = 30°$, according to experimental observation (and reference CA-based simulation), with all other Euler angles being

$\varphi = \theta = 0°$. The boundary conditions are chosen to result in a constant thermal gradient and cooling rate, and a pulling velocity of $v_{pull} = 86$ µm/s. Following [30], the kinetic model was set as Eq. (2) with $a_1 = 8.26 \times 10^{-6}$ m s$^{-1}$ K$^{-2}$, $b_1 = 2$, $a_2 = 8.18 \times 10^{-5}$ m s$^{-1}$ K$^{-3}$, and $b_2 = 3$. Further details on the domain, discretization, thermophysical properties, initial and boundary conditions are given in Table S2 of the joint Supplementary Material.

Focusing on the right-hand-side GB between the central grain at $\psi = 30°$ and the right grain with $\psi = 11°$, Figure 5 compares the precited GB angle by the current model and the CA results from [30] as a function of the microscopic growth model spatial discretization (i.e. $\lambda_0$ in the present model, and the CA grid element size in [30]). The convergence trends exhibited by the CA and the particle-based approach are similar, with the current model overall resulting in a GB angle closer to the fully converged result, which is close to the experimentally measured angle $\approx 19°$. The present model appears reasonably converged for any $\lambda_0 \leq 48$ µm, which is higher than the experimentally measured SDAS $\lambda_2 = 8.4 \pm 0.4$ µm [61]. It should be noted that, the uncertainty on the measurement of the GB angle is of order $\pm 1°$ to $\pm 2°$ for coarser grids ($\lambda_0 > 55$ µm), while it is a vanishingly small fraction of a degree for finer grids ($\lambda_0 < 55$ µm). (This is mostly due to the selection of exact locations for GB start and end points between laterally adjacent yet increasingly far-apart particles.)

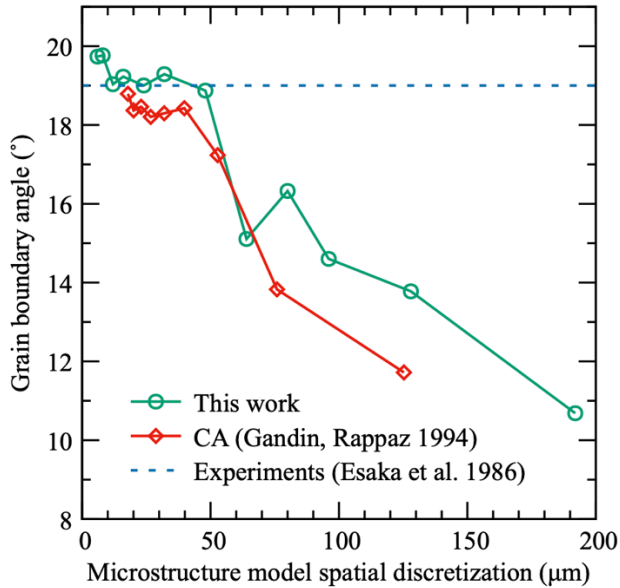

Figure 5: Average orientation of the grain boundary between grains titled by 30° and 11° predicted by the current particle-based model compared to published CA-FE results [30] and experimental results by Esaka et al. [61].

Even though the optimal choice of cell size in CA (and, by extension, the choice of $\lambda_0$ in our particle-based approach) in the context of grain growth competition is still debated [37][62][38], these results suggest that similar convergence trends might be expected from the present model. In the next subsections, we move on to more complex 3D polycrystalline cases, without (Section 3.4) and with (Section 3.5) nucleation.

### 3.4. Directional solidification of three-dimensional polycrystal

In the next test of our model and implementation, we simulate the polycrystalline competitive growth of dendritic grains in a three dimensional (3D) configuration, and compare our results with phase-field results published by Takaki et al. [63]. The simulations correspond to the directional solidification of an Al-3 wt% Cu alloy in the $z$ direction at a velocity $v_{pull} = 100$ µm/s under various temperature gradients and initial grain distributions (referred to as cases 1, 2, and 3 in [63]). Here, the temperature field is imposed using the classical "frozen temperature approximation" (i.e. using homogeneous and constant temperature gradient and cooling rate), and the diffusive solute concentration field is coupled to the particle-based solidification solver. We focus on cases 1 and 3 for $G = 10$ K/mm and 100 K/mm. Periodic boundary conditions were applied on domain boundaries normal to the $x$ and $y$ directions. For the growth kinetics, we used Eq. (1) with $b = 2.282$ and $a = 1.834 \times 10^{-4}$ m s$^{-1}$ K$^{-b}$. The spatial distribution and grain orientation of the initial nuclei for cases 1 and 3 are as given in the Supplementary Materials of Ref. [63]. Further details on input parameters including spatial and temporal discretization and thermophysical properties are provided in Table S3 of the attached Supplementary Materials.

Figure 6 shows a top-view (i.e. with the main growth direction, $z$, oriented towards the reader) of the solidification front, colored by the angle between the $z$-axis and the [100] crystalline direction, at three different time steps $t \approx 8$, 27 and 133 s ($\pm 1$ s), for each of the four cases, comparing our results to published PF results [63] using the same color palette. While the results of the two models do not exhibit an exact match, the overall trends are qualitatively well reproduced, in terms of grain elimination rates (Figure 7).

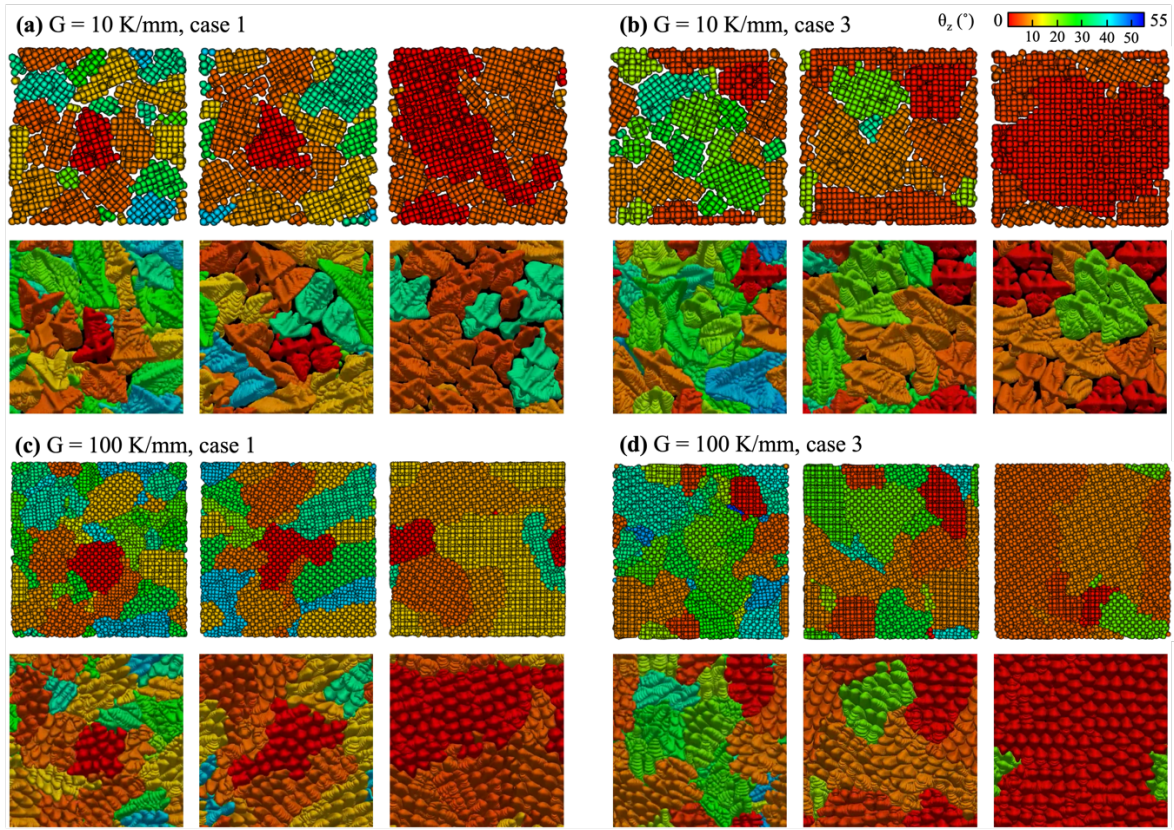

*Figure 6: Top view of the solid-liquid interface for a temperature gradient $G = 10$ K/mm (a,b) and 100 K/mm (c,d) for initial grain distributions corresponding to case 1 (a,c) and 3 (b,d). For each case, we compare the results of the particle-based approach (top) and phase field model [63] (bottom), for $t \approx 8$, 27 and 133 s ($\pm 1$ s)(left to right). Lateral dimensions are both 768 μm.*

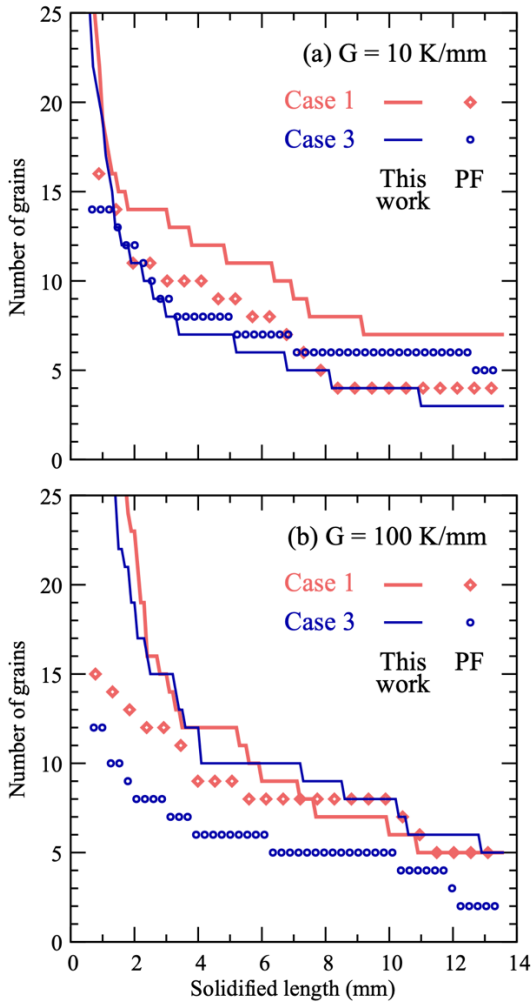

Figure 7: Number of surviving grains as a function of the solidified length for cases 1 and 3 predicted by the particle-based model (lines), compared to PF result by Takaki et al. [63] (symbols) for a thermal gradient of G = 10 K/mm (a) and 100 K/mm (b).

Indeed, even though their precise spatial distribution is not exactly similar, results of the particle-based approach show similar color shades within proportions comparable to PF results. As already discussed in previous articles (e.g. [38]), discrepancies are expected between PF and coarse-grained envelope-based approaches. PF results tend to be stochastic in nature, due to the fact that dendritic side-branches have their origin in the selective amplification of fluctuations. (Whether or not PF simulations are performed with noise, minor changes in initial conditions or numerical resolution may lead to noticeable differences at diverging GBs.) Diverging GBs thus exhibit a high lateral mobility since their morphology results from side-branching competition [59][60]. The use of growth kinetics based on steady-state assumptions (e.g. Ivantsov) also leads to discrepancies with PF results that numerically resolve the transient diffusion of species ahead of each dendrite tip. Another source of discrepancy comes from the fact that dendrites do not always grow perfectly aligned with their <100> directions [59],[66]-[68], while they are assumed to do so in coarse-grained models. This is particularly important for higher temperature gradient, which may lead to an intermediate cellular-dendritic regime where the cells/dendrites grow more closely aligned to the temperature gradient direction, hence resulting in significantly different growth competition mechanisms [63][69].

In these specific simulations, the temperature field is imposed, and the grain growth competition occurs through the solute field, just like in PF simulations, as the local undercooling entering the growth kinetics is estimated with respect to the local solute concentration. This allows calculating the solute concentration map within the solidified domain, as illustrated in Figure 8 for the two simulations with G = 10 K/mm. The cross sections (a,b) and longitudinal sections (c,d) of the solute concentration field clearly highlight interdendritic segregation patterns within the grains, as well as along grain boundaries. Still, it is worth noting that, in order to obtain this level of detail, both the grid element size $\Delta x$ and the microstructural length scale $\lambda_0$ need to be taken relatively small compared to the primary dendrite arm spacing, which may be inconvenient for computational purposes.

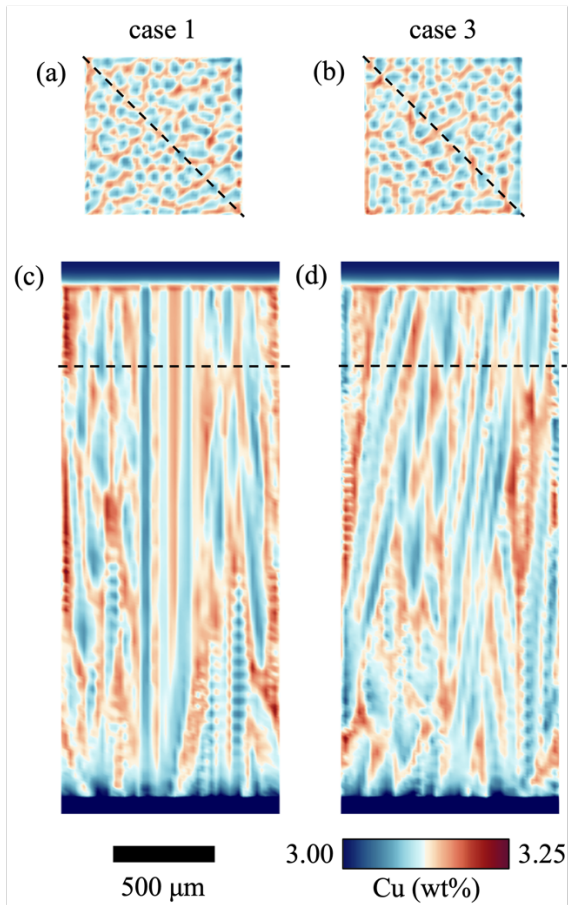

*Figure 8: Solute concentration maps for the two cases at G = 10 K/mm, within a cross section (a,b) and a longitudinal section (c,d) marked with dashed lines, pictured at t = 27 s.*

In spite of the inability of coarse-grained envelope-based models to reproduce fine details of grain growth competition, it was previously shown that, when focusing on average distributions over a statistically representative number of grains, a reasonable agreement with PF results can be achieved if the microstructural model discretization is fine enough [38]. Here, the total number of grains is not sufficient to extract meaningful orientation distribution histograms. However, a quantitative and averaged comparison can still be drawn by comparing the number of grains as a function of the grown length, as in Figure 7. These plots show that, despite some expected deviation, the predicted grain coarsening/elimination rate is close to PF results. The elimination rate from the particle-based model appears slightly slower (i.e. higher number of grains), but not by a significant amount. Figure 7 also shows the increasing discrepancy expected with a higher temperature gradient due to the transition toward an intermediate cellular-dendritic regime. These results are satisfactory, given that the benchmark is applied with a much more accurate yet much more demanding phase-field model.

In terms of computational cost, our model can afford a FV spatial discretization step of 16 μm with a microstructural scale $\lambda_0$ = 24 μm (for G = 10 K/mm) or 10 μm (100 K/mm), compared to the fine grid element size of 0.75 μm required for well-converged PF simulations. Each simulation was performed in between 15.7 h (for $G$ = 10 K/mm) and 22.5 h ($G$ = 100 K/mm) of wall clock time using 24 CPU cores of an AMD EPYC 7713 Processor, while the PF model requires about 5 days on 144 GPUs [63].

### 3.5. Columnar-to-equiaxed transition in an Al-Si alloy

As a final verification, we simulate the solidification and resulting columnar-to-equiaxed transition in a cylindrical Al-7wt%Si ingot, corresponding to experiments by Gandin [70]. The problem involves a diffusive thermal field coupled with the particle-based solidification solver (no solute field). While the grain distribution and orientation are randomized, initial and boundary conditions are taken according to subsequent CA-FE simulations performed by Carozzani et al. [33]. Nuclei of uniform random orientation distribution and location are initialized across the cooled lower domain boundary and within the volume at a given nuclei density. The nucleation undercooling of the boundary nuclei is set to zero, whereas it exhibits a normal distribution for nuclei in the bulk with $\Delta T_{min}$ = 0 K, $\Delta T_{max}$ = 10 K, $\Delta T_\mu$ = 5.19 K, and $\Delta T_\sigma$ = 0.25 K. The thermal boundary condition at the lower domain boundary is identical to Ref. [33], where the experimentally obtained temperature values of [70] are extrapolated to $z$ = 0 (see Figure 10). Following [33], a heat flux of 3000 W/m$^2$ is applied to the top boundary for t ≤ 900 s, in order to account for heat losses at the liquid-gas interface resulting from the air gap forming due to solidification shrinkage. The full list of material parameters, initial and boundary conditions and numerical parameters used in this simulation are provided in Table S4 of the attached Supplementary Materials, with temperature-dependent density and thermal conductivity listed in Tables S5 and S6.

Figure 9 shows the resulting evolution of the grain structure, and Figure 10 shows the evolution of temperature in the ingot measured at different heights $z$ within the ingot (measured along its central axis). The predicted grain structure exhibits a clear columnar-to-equiaxed transition (CET), with elongated columnar grains at the bottom and equiaxed grains at the top, which is consistent with experiments [70] and CA simulations [33]. The CET occurs progressively between approximately 100 and 110 mm in height, compared to ≈ 118 mm in the experiment and ≈ 108 mm in the CA-based simulation. The temperature evolution at the $z$ = 140 mm probe clearly exhibits a recalescence just below the alloy liquidus temperature (≈ 618°C), which can only be reproduced using a model accounting for undercooled nucleation and grain growth. Like in Ref. [33], discrepancies between experiments and simulations can be attributed to several effects that are not included in the model, such as sedimentation, fragmentation, shrinkage, and macrosegregation driven by fluid flow. The CA-FE simulation in [33] required a CA cell size of 250 μm resulting in 43×10$^6$ cells. In comparison, the number of active tracking points in the present particle-based approach resulted in a peak in active particles of 413 766 at t = 906 s, with an average number of about 206 000 active particles over the entire simulation. The main conclusion from these results (like those previously presented in this section) is that the current approach allows simulations with a similar accuracy as reference grain-scale simulations methods, such as CA, yet with a more data- and computation-efficient particle-based approach.

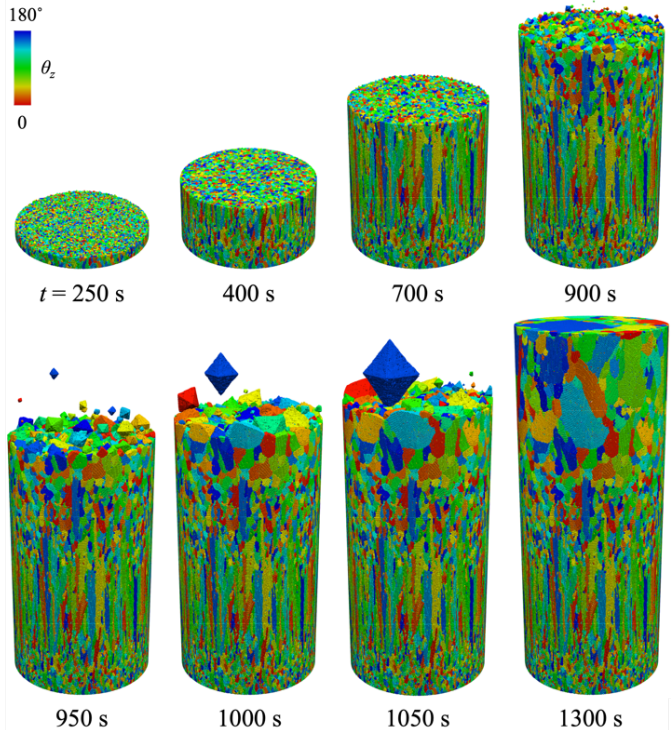

*Figure 9: Grain structure evolution during 3D simulation of columnar-to-equiaxed transition in Al-7wt.%Si ingot of height 173 mm and diameter 70 mm.*

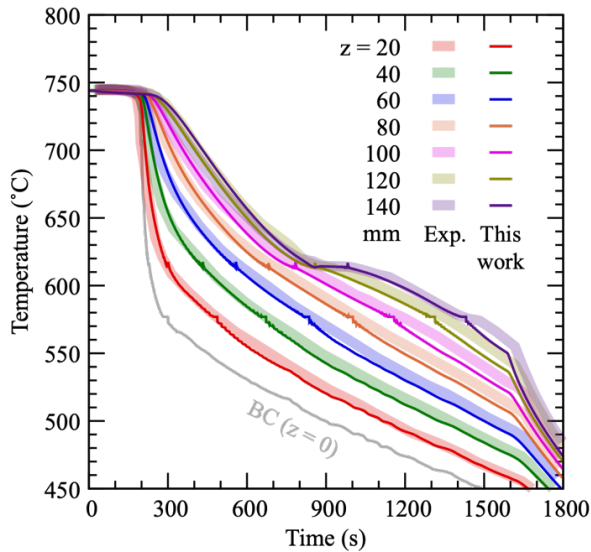

*Figure 10: Temperature evolution at different locations in the ingot, comparing experimental measurements from [70] (thick pale lines) with predictions of the FV model coupled with the particle-based grain nucleation growth (thin darker lines), also showing the imposed BC at the lower end (z=0) of the ingot (gray line).*

## 4. Original applications

### 4.1. Melting test case

Before applying our solver to a laser welding case (Section 4.2), we test the implementation of melting described in Section 2.1.3. To do so, we set up a simple, quasi-two-dimensional case, where the initial random Voronoi-based grain structure is partially melted and re-solidified under a steep thermal gradient relevant to industrial processes, such as laser-beam melting.

The considered quasi-2D domain (with dimensions $L_x \gg L_z$ and $L_y \gg L_z$) is sketched in the inset of Figure 11. Two different Dirichlet boundary conditions (BCs) are used, denoted as *isothermal* on the bottom (y-) and right (x+) boundaries and *constrained* on the top (y+) and left (x-) boundaries. Boundary conditions imposed on the front and back boundaries along the short $z$ direction are Neumann (no-flux) BCs (i.e. so-called "empty" BCs in OpenFOAM). The *constrained* boundary serves as simplified model for a high-energy beam (e.g., laser) melting process, i.e. with a steep temperature increase followed by a slower decrease in temperature. The imposed Dirichlet BCs are plotted in Figure 11. Initially, nuclei are seeded with a density of $\rho_{nuc} = 10^{13}$ m$^{-3}$ throughout the domain, and random three-dimensional ($\psi, \theta, \varphi$) orientation. From these nuclei, an initial grain structure develops from the cooling of the domain below the liquidus temperature in the first 2.5 ms of the simulation. Then, a steep rise in temperature at the *constrained* boundary serves as simple model for a high-energy beam heat input, followed by a linear cooldown at constant cooling rate, $dT/dt = 2 \times 10^4$ K/s, representative of laser-based processing. For the growth kinetics model, we use Eq. (2) with coefficients $a_1 = 8.315 \times 10^{-6}$ m s$^{-1}$ K$^{-2.49}$, $b_1 = 2.49$, $a_2 = 9.628 \times 10^{-7}$ m s$^{-1}$ K$^{-3.622}$, and $b_2 = 3.622$, which corresponds to a stainless steel [35]. All case-specific parameters, such as thermophysical properties, initial conditions, and domain size, are summarized in Table S7 of the joint supplementary material.

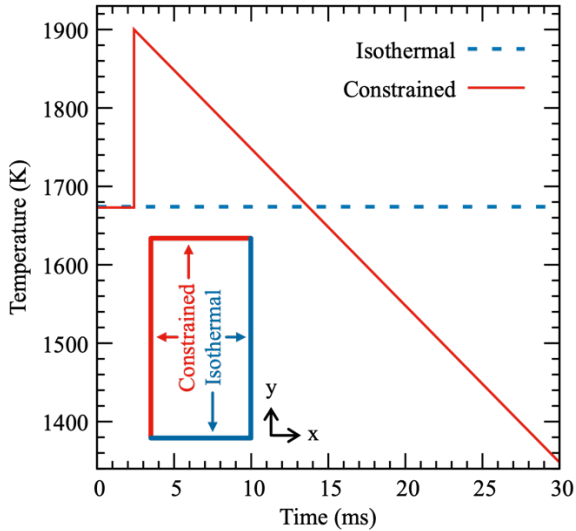

Figure 11: Temperature evolution profiles imposed as Dirichlet conditions on the x- and y-direction boundaries. The inset schematics illustrates the location of "isothermal" and "constrained" conditions in the (x,y) plane.

The resulting grain structure evolution is illustrated in Figure 12, where the grains are colored by their [100] orientation with respect to the $\vec{e}_y$-axis. These results show that the considered approach and algorithm for melting (Section 2.1.3), i.e. tracing back the particles up to the nominal liquidus temperature, qualitatively reproduces the expected behavior for a melted and (re)solidified structure.

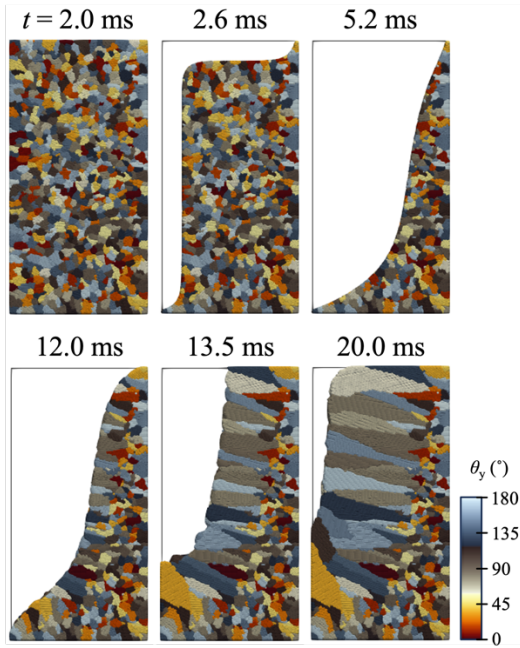

*Figure 12: Grain structure at different time steps of the melting and solidification simulation, with grains colored by their crystal orientation within the (x,y) plane, $\theta_y$. (Since grains have random three-dimensional orientations, the chosen color scale is mostly illustrative.)*

### 4.2. Laser beam welding of 316L stainless steel

Now that the solidification and melting model and its implementation have been validated, we illustrate an industrially relevant application. At the same time, we also illustrate that the particle-based approach can be coupled with nearly any custom CFD solver. To do so, we focus on keyhole-mode laser beam welding (LBW) of a stainless steel. The particle-based model (referred to as the *solidification* solver) was coupled with a multiphysics laser-based manufacturing solver (referred to as the *CFD* solver) [71].

The OpenFOAM-based CFD solver was described in detail in [55]. Its capability for predicting real-world processes was demonstrated for a broad range of processes, such as the humping phenomenon and transition to cutting in welding of steel [72], dissimilar metal welding [73], overlap welding with beam shaping under the presence of an interfacial gap [74], or additive manufacturing via powder bed fusion [75]. A detailed description of the solver, its underlying mathematical framework and numerical implementation, including multiple validation cases including laser beam welding, was recently published elsewhere [55].

While in Section 3, the particle-based solver and transport FV solver are coupled in both directions (even though not iteratively), here we use a simplified coupling scheme. The coupling between the CFD and solidification solvers is of a one-way nature (weak coupling): the solidification solver receives information on the distribution of phases (solid, liquid, and gaseous metal, as well as ambient gas) and temperature distribution at each CFD simulation time step to calculate the resulting grain structure evolution. A two-way coupling, with information feedback from the solidification solver to the CFD solver is also feasible, but here we simply illustrate that the approach can also be applied as a post-processing tool to any complex multiphysics solver.

In welding, the solid base metal acts as seed for epitaxial grain growth. In spite of some variability, the resulting grain growth competition typically favors grains oriented normal to the liquid-solid interface, i.e. along the main temperature gradient direction [60][38]. Keyhole mode welding, as opposed to conduction mode welding, is characterized by high temperatures, locally reaching the boiling point. This leads to evaporation, with the recoil pressure forming a capillary, which eventually reaches a keyhole shape, in which the laser beam is reflected multiple times, increasing the amount of laser power further absorbed by the material. The resulting melt pool and final weld seam have a high depth-to-width ratio [76]. The microstructure obtained through laser beam welding is typically determined by the high peak temperatures, a moving solidification front, and negligible constitutional and kinetic undercoolings [77].

The simulated configuration features a 2 mm sheet of 316L stainless steel welded in total penetration mode, as investigated experimentally and analytically by Artinov et al. [78]. The material properties of 316L stainless steel are set using the values reported in [79] and [80]. An initial structure is established in the base material with randomly distributed seeds (see Figure 13). The initial grain structure is only placed within a small portion of the entire domain, where the process is expected to be in steady-state mode, to avoid computational costs associated with computing the grain structure within the entire domain. This initial structure is then locally "erased" by melting during the welding process (see Section 4.1), allowing for re-growth under the thermal gradients and cooling rates determined by the CFD simulation. The computational domain, FV mesh and initially generated grain structure are illustrated in Figure 13. Full details on the initialization parameters, kinetic model, process and numerical parameters are provided in Table S8, with temperature-dependent material properties listed in Table S9 of the Supplementary Material.

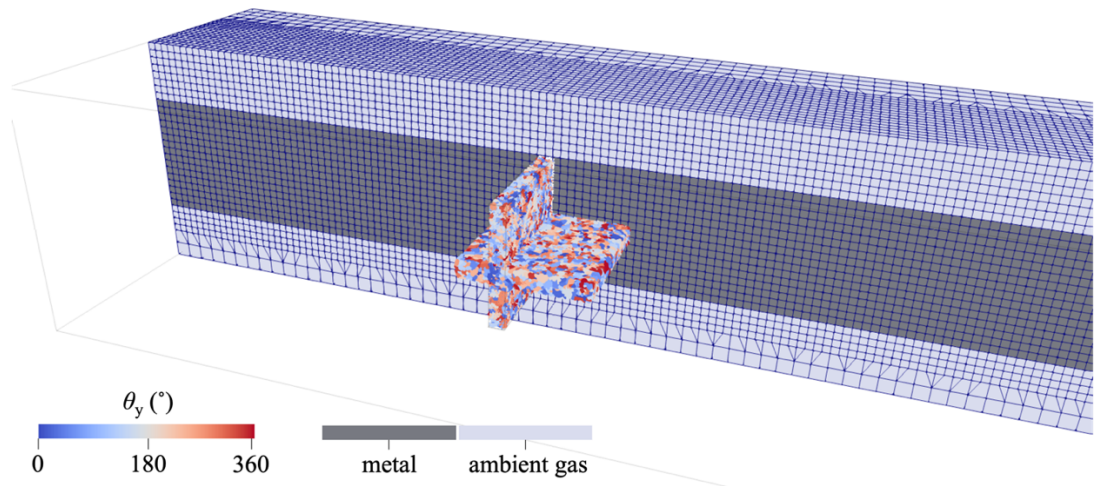

*Figure 13: Computational domain showing mesh and initial phase distribution (cut along longitudinal section, showing only half of domain), and initial grain structure generated from randomly seeded nuclei, placed in a horizontal and transverse section.*

Figure 14 shows an image of the simulated melt pool and surrounding temperature field once it has reached a steady state (i.e. when the melt pool length is not changing anymore). The resulting grain structure evolution is illustrated in horizontal and transverse sections in Figure 15. As expected, a structure develops with elongated grains roughly oriented normal to the solidification front. The horizontal section also shows a slight tendency of favoring grains oriented following the welding direction, due to the shape of the melt pool tail. The solid-liquid interface formed by the grain envelopes remains located within the mushy zone (i.e. the region with a liquid fraction between zero and one).

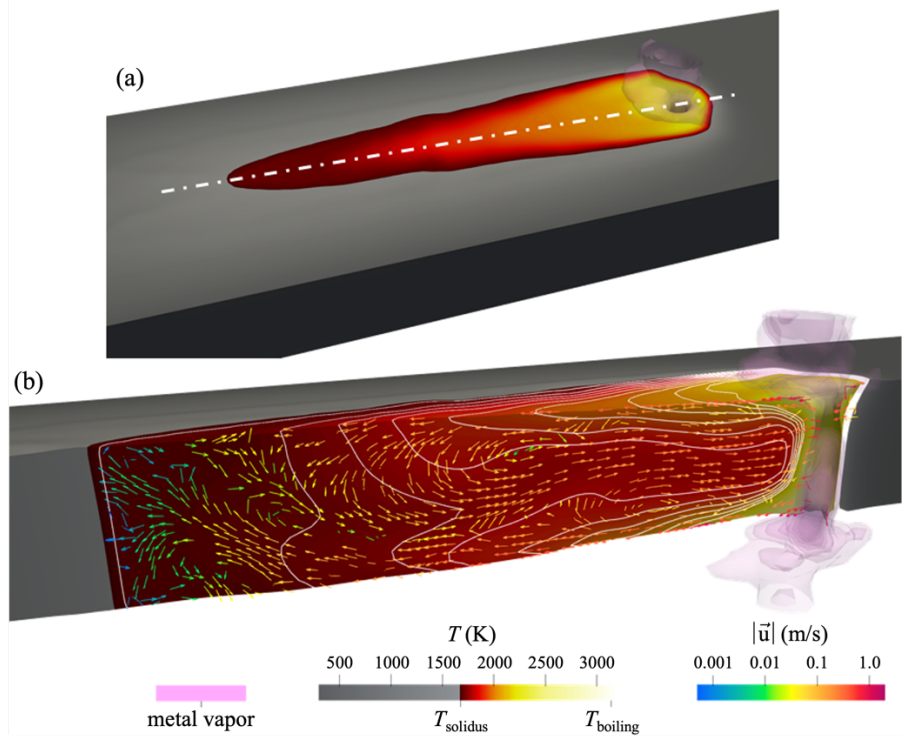

*Figure 14: Top view (a) and longitudinal section (b) along the laser path (white dash-dotted line) during the welding simulation, showing temperature field as color map with contour lines at steps of 50 K between 1700 K and 2100 K, as well melt flow (arrows) and evaporated metal (translucent pink shading).*

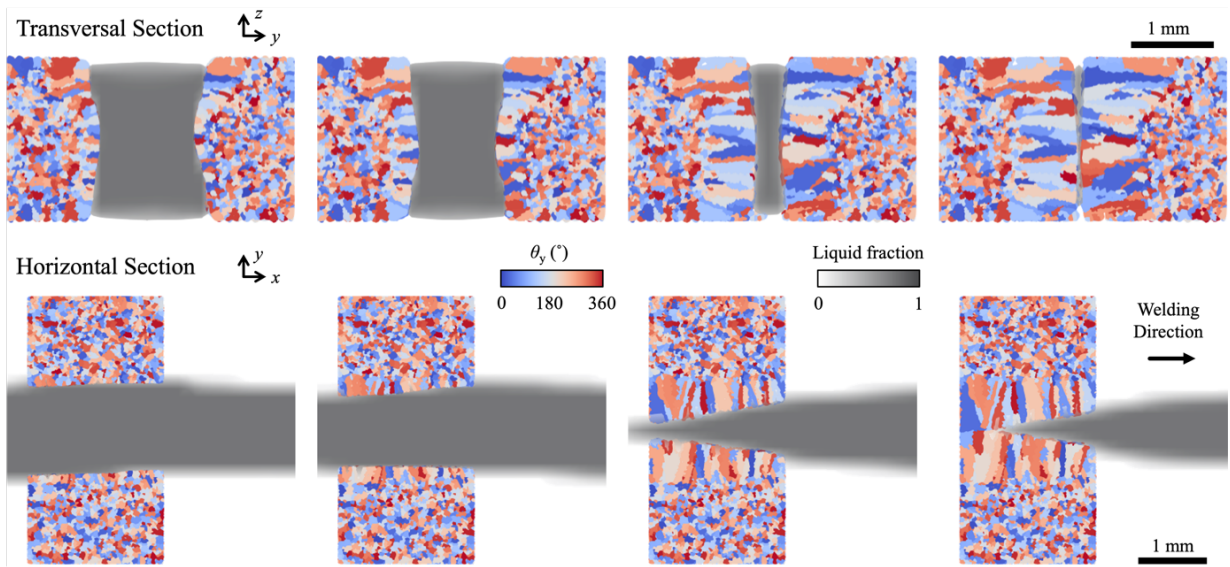

*Figure 15: Transversal and horizontal sections at different times during solidification, showing grains (colored by orientation) and liquid metal (grayscale).*

Although qualitative, these results illustrate that our particle-based solidification solver can be integrated within any CFD solver. The entire coupled simulation took approximately 80h to complete on 8 CPU cores of a standard desktop computer. Notably, during the peak in computational load (competitive re-growth in solidifying melt pool), nearly 25% of overall CPU time was consumed by the grain growth tracking algorithm, therefore not standing out as a severe bottleneck even in a non-optimized "proof-of-concept" simulation that did not rely on GPU acceleration.

## 5. Summary and outlook

We presented an original approach for the modeling of solidification applicable at the scale of three-dimensional grain structures (i.e. microstructure representative volume elements). In terms of underlying physics and assumptions, the model relies on classical equations common to state-of-the-art "mesoscale" models, in particular those considered for cellular automaton models.

The key novelty is the use of a grain envelope tracking scheme using a Lagrangian point-based discretization. The evolution of the tracking points is inspired by the self-similar hierarchical pattern of dendritic crystals. The computational task of nearest neighbor search for grain "collision" detection between particles in competitive growth scenarios (i.e. the formation of grain boundaries) is implemented in a way that leverages CPU/GPU-parallelization using C++ libraries, e.g. OpenVDB. The VDB data structure is hierarchically designed and provides $\mathcal{O}(1)$ random stencil data access [57], thus making it well suited for nearest neighbor search and collision detection of narrow band level sets (computationally challenging in competitive grain growth scenarios). Hence, we expect both one-way and two-way coupled solvers to be highly scalable [57][58][81], which will be fully addressed in a follow-up article. Hence, the approach is suitable for both small-scale scenarios encountered in laser-based manufacturing as well as large-scale scenarios such as ingot casting.

We validated the model against various benchmark cases, including single crystal growth where an analytical solution is available, two- and three-dimensional competitive grain growth scenarios, as well as columnar-to-equiaxed transition in three dimensions. Agreement with analytical, experimental, and numerical (CA) solutions is satisfactory. Finally, we illustrated the potential of the approach to be coupled with a multiphysics laser-processing simulation model. This last example showcases the straightforward inclusion of three-dimensional competitive grain growth in a thermo-fluid-mechanical model. It should be noted that the results of the particle-based approach – like these of a CA model – are strongly dependent on the choice of the underlying volume-averaged model, and on its appropriateness to tackle the problem at hand.

Ongoing and future work includes a thorough study of the approach scalability including massive parallelization of the tracking algorithm on large GPU architectures, which has shown promising results but remains to be quantified in a more systematic basis [58]. The most important extensions foreseen include the incorporation of the movement of solid grains, and/or solid-state microstructure evolution (e.g. grain growth or formation of secondary phases in the heat affected zone adjacent to a laser path). We expect the present approach to constitute a valuable alternative to existing models, hence applicable to large scale relevant to industrial cases (e.g., industrial parts manufactured by laser-based powder bed fusion).

**CRediT authorship contribution statement**

**Salem Mosbah**: Conceptualization, Methodology, Software, Formal analysis, Resources, Supervision, Writing – Original Draft. **Rodrigo Gómez Vázquez**: Methodology, Software, Validation, Formal analysis, Investigation, Visualization, Writing – Original Draft. **Constantin Zenz**: Methodology, Software, Validation, Formal analysis, Investigation, Visualization, Writing – Original Draft. **Damien Tourret**: Formal analysis, Visualization, Writing – Review & Editing. **Andreas Otto**: Supervision, Resources, Funding acquisition, Writing – Review & Editing.

**Data availability**

Data will be made available on request.

**Acknowledgements**

The authors wish to thank Charles-André Gandin and Tomohiro Takaki for kindly providing data of their original simulations to compare with our results. This work was funded in part by the European Union Horizon 2020 research and innovation program under grant agreement number 825103. CZ was supported by TU Wien Doctoral School, within the Doctoral College DigiPhot. DT gratefully acknowledges support from the Spanish Ministry of Science through a Ramón y Cajal Fellowship (Ref. RYC2019-028233-I).
**References**

[1] J. Allison, D. Backman, L. Christodoulou. *Integrated computational materials engineering: a new paradigm for the global materials profession*. JOM 58 (2006) 25-27.

[2] J. Panchal, S.R. Kalidindi, D.L. McDowell. *Key computational modeling issues in integrated computational materials engineering*. Computer-Aided Design 45 (2013) 4-25.

[3] W.Y. Wang, J. Li, W. Liu, Z.K. Liu. *Integrated computational materials engineering for advanced materials: A brief review*. Computational Materials Science 158 (2019) 42-48.

[4] W. Xiong, G. Olson. *Cybermaterials: materials by design and accelerated insertion of materials*. npj Computational Materials 2 (2016) 15009.

[5] A. Karma, D. Tourret. *Atomistic to continuum modeling of solidification microstructures*. Current Opinion in Solid State and Materials Science 20 (2016) 25-36.

[6] M. Rappaz. *Modeling and characterization of grain structures and defects in solidification*. Current Opinion in Solid State and Materials Science 20 (2016) 37-45.

[7] W. Shao, J.M. Guevara-Vela, A. Fernández-Caballero, S. Liu, and J. LLorca. *Accurate prediction of the solid-state region of the Ni-Al phase diagram including configurational and vibrational entropy and magnetic effects*. Acta Materialia 253 (2023) 118962.

[8] J.J. Hoyt, M. Asta, and A. Karma. *Atomistic and continuum modeling of dendritic solidification*. Materials Science and Engineering: R 41 (2003) 121-163.

[9] W.J. Boettinger, J.A. Warren, C. Beckermann, A. Karma. *Phase-field simulation of solidification*. Annual review of materials research 32 (2002) 163-194.

[10] N. Provatas, K. Elder. *Phase-field methods in materials science and engineering*. John Wiley & Sons (2011).

[11] D. Tourret, H. Liu, J. LLorca. *Phase-field modeling of microstructure evolution: Recent applications, perspectives and challenges*. Progress in Materials Science 123 (2022) 100810.

[12] T. Shimokawabe, T. Aoki, T. Takaki, T. Endo, A. Yamanaka, N. Maruyama, A. Nukada, S. Matsuoka. *Peta-scale phase-field simulation for dendritic solidification on the TSUBAME 2.0 supercomputer*. In: Proceedings of 2011 International Conference for High Performance Computing, Networking, Storage and Analysis (2011) 1-11

[13] J. Hötzer, A. Reiter, H. Hierl, P. Steinmetz, M. Selzer, B. Nestler. *The parallel multi-physics phase-field framework Pace3D*. Journal of computational science 26 (2018) 1-12.

[14] S. Sakane, T. Takaki, T. Aoki. *Parallel-GPU-accelerated adaptive mesh refinement for three-dimensional phase-field simulation of dendritic growth during solidification of binary alloy*. Materials Theory 6 (2022) 3.

[15] M. Greenwood, K.N. Shampur, N. Ofori-Opoku, T. Pinomaa, L. Wang, S. Gurevich, N. Provatas. *Quantitative 3D phase field modelling of solidification using next-generation adaptive mesh refinement*. Computational Materials Science 142 (2018) 153-171.

[16] W. Feng, P. Yu, S. Hu, Z. Liu, Q. Du, L. Chen. *Spectral implementation of an adaptive moving mesh method for phase-field equations*. J. Comput. Phys. 220 (2006) 498–510.

[17] L. Chen, J. Shen. *Applications of semi-implicit Fourier-spectral method to phase field equations, Comput*. Phys. Comm. 108 (1998) 147–158.

[18] J. Zhu, L.-Q. Chen, J. Shen, V. Tikare. *Coarsening kinetics from a variable mobility Cahn–Hilliard equation: Application of a semi-implicit Fourier spectral method*. Phys. Rev. E 60 (1999) 3564.

[19] A.D. Boccardo, M. Tong, S.B. Leen, D. Tourret, J. Segurado. *Efficiency and accuracy of GPU-parallelized Fourier spectral methods for solving phase-field models*. Computational Materials Science 228 (2023) 112313.

[20] J. Ni, C. Beckermann. *A volume-averaged two-phase model for transport phenomena during solidification*. Metallurgical Transactions B 22 (1991) 349-361.

[21] C.Y. Wang, C. Beckermann. *A multiphase solute diffusion model for dendritic alloy solidification*. Metallurgical and Materials Transactions A 24 (1993) 2787-2802.

[22] H. Combeau, M. Založnik, S. Hans, P.E. Richy. *Prediction of macrosegregation in steel ingots: Influence of the motion and the morphology of equiaxed grains*. Metallurgical and materials transactions B 40 (2009) 289-304.

[23] J. Li, M. Wu, A. Ludwig, A. Kharicha. *Simulation of macrosegregation in a 2.45-ton steel ingot using a three-phase mixed columnar-equiaxed model*. International journal of heat and mass transfer 72 (2014) 668-679.

[24] D. Tourret, A. Karma. *Three-dimensional dendritic needle network model for alloy solidification*. Acta Materialia 120 (2016) 240-254.

[25] T. Isensee, D. Tourret. *Convective effects on columnar dendritic solidification–A multiscale dendritic needle network study*. Acta Materialia 234 (2022) 118035.

[26] I. Steinbach, C. Beckermann, B. Kauerauf, Q. Li, J. Guo. *Three-dimensional modeling of equiaxed dendritic growth on a mesoscopic scale*. Acta materialia 47 (1999) 971-982.

[27] Y. Souhar, V.F. De Felice, C. Beckermann, H. Combeau, M. Založnik. *Three-dimensional mesoscopic modeling of equiaxed dendritic solidification of a binary alloy*. Computational Materials Science 112 (2016) 304-317.

[28] T.M. Rodgers, J.D. Madison, V. Tikare. *Simulation of metal additive manufacturing microstructures using kinetic Monte Carlo*. Computational Materials Science 135 (2017) 78-89.

[29] A.F. Chadwick, P.W. Voorhees. *The development of grain structure during additive manufacturing*. Acta Materialia 211 (2021) 116862.

[30] Ch.-A. Gandin and M. Rappaz. *A coupled finite element-cellular automaton model for the prediction of dendritic grain structures in solidification processes*. Acta Metallurgica et Materialia, 42 (1994) 2233-2246.

[31] Ch.-A. Gandin and M. Rappaz. *A 3D Cellular Automaton algorithm for the prediction of dendritic grain growth*. Acta Materialia 45 (1997) 2187-2195.

[32] H. Takatani, Ch.-A. Gandin, M. Rappaz. *EBSD Characterisation and Modeling of Columnar Dendritic Grains Growing in the Presence of Fluid Flow*. Acta Materialia 48 (2000) 675-688.

[33] T. Carozzani, H. Digonnet, Ch.-A. Gandin. *3D CAFE modeling of grain structures: application to primary dendritic and secondary eutectic solidification*. Modelling Simul. Mater. Sci. Eng. 20 (2012) 015010.

[34] T. Carozzani, Ch.-A. Gandin, H. Digonnet, M. Bellet, K. Zaidat, Y. Fautrelle. *Direct simulation of a solidification benchmark experiment*. Metallurgical and Materials Transactions A 44 (2013) 873-887.

[35] S. Chen, G. Guillemot, Ch.-A. Gandin. *Three-dimensional cellular automaton-finite element modeling of solidification grain structures for arc-welding processes*. Acta materialia 115 (2016) 448-467.

[36] A. Rai, M. Markl, C. Körner. *A coupled Cellular Automaton–Lattice Boltzmann model for grain structure simulation during additive manufacturing*. Computational Materials Science 124 (2016) 37-48.

[37] A. Pineau, G. Guillemot, D. Tourret, A. Karma, and Ch-A. Gandin. *Growth competition between columnar dendritic grains–Cellular automaton versus phase field modeling*. Acta Materialia 155 (2018) 286-301.

[38] S.M. Elahi, R. Tavakoli, I. Romero, and D. Tourret. *Grain growth competition during melt pool solidification — Comparing phase-field and cellular automaton models*. Computational Materials Science 216 (2023) 111882.

[39] J.A. Dantzig and M. Rappaz. *Solidification*, EFPL Press (2009).

[40] W. Kurz, D. Fisher and M. Rappaz. *Fundamentals of solidification* (5th edition), EFPL Press (2023).

[41] L. Gránásy, G.I. Tóth, J.A. Warren, F. Podmaniczky, G. Tegze, L. Rátkai, and T. Pusztai. *Phase-field modeling of crystal nucleation in undercooled liquids – A review*. Progress in Materials Science 106 (2019) 100569.

[42] W. Kurz, B. Giovanola, and R. Trivedi. *Theory of microstructural development during rapid solidification*. Acta Metallurgica 34 (1986) 823-830.

[43] M. Rappaz, S.A. David, J.M. Vitek, and L.A. Boatner. *Analysis of solidification microstructures in Fe-Ni-Cr single-crystal welds*. Metallurgical Transactions A 21 (1990) 1767-1782.

[44] G. P. Ivantsov. *Temperature field around a spherical, cylindrical, and needle-shaped crystal, growing in a pre-cooled melt*. Doklady, Akademiya Nauk SSR 58 (1947) 567-569.

[45] W. Kurz, M. Rappaz, and R. Trivedi. *Progress in modelling solidification microstructures in metals and alloys. Part II: dendrites from 2001 to 2018*. International Materials Reviews 66 (2021) 30-76.

[46] B. Cantor, and A. Vogel. *Dendritic solidification and fluid flow*. Journal of Crystal Growth 41 (1977) 109-123.

[47] J.S. Langer. *Instabilities and pattern formation in crystal growth*. Reviews of modern physics 52 (1980) 1.

[48] A. Barbieri, and J. S. Langer. *Predictions of dendritic growth rates in the linearized solvability theory*. Physical Review A 39 (1989) 5314.

[49] M. Bobadilla, J. Lacaze and G. Lesoult. *Influence des conditions de solidification sur le déroulement de la solidification des aciers inoxydables austénitiques*. J. Cryst. Growth, 89 (1988) 531-544

[50] M. Rappaz, and W. Boettinger. *On dendritic solidification of multicomponent alloys with unequal liquid diffusion coefficients*. Acta Materialia, 47 (1999) 3205-3219.

[51] H. Combeau, M. Bellet, Y. Fautrelle, D. Gobin, E. Arquis, O. Budenkova, B. Dussoubs, et al. *Analysis of a numerical benchmark for columnar solidification of binary alloys*. IOP Conference Series: Materials Science and Engineering 33 (2012) 012086.

[52] V.R. Voller, C. Prakash. *A fixed grid numerical modelling methodology for convection-diffusion mushy region phase-change problems*. International Journal of Heat and Mass Transfer 30 (1987) 1709-1719.

[53] Ch-A. Gandin, S. Mosbah, T. Volkmann, and D.M. Herlach. *Experimental and numerical modeling of equiaxed solidification in metallic alloys*. Acta Materialia 56 (2008) 3023-3035.

[54] D. Tourret, Ch-A. Gandin, T. Volkmann, and D.M. Herlach. *Multiple non-equilibrium phase transformations: Modeling versus electro-magnetic levitation experiment*. Acta materialia 59 (2011) 4665-4677.

[55] C. Zenz, M. Buttazzoni, T. Florian, K.E. Crespo Armijos, R. Gómez Vázquez, G. Liedl, and A. Otto. *A compressible multiphase mass-of-fluid model for the simulation of laser-based manufacturing processes*. Computers & Fluids 268 (2024) 106109.

[56] H. Jasak, A. Jemcov, and Z. Tukovic. *OpenFOAM: A C++ library for complex physics simulations*. In: International workshop on coupled methods in numerical dynamics, vol. 1000 (2007) 1-20.

[57] K. Museth. *VDB: High-Resolution Sparse Volumes with Dynamic Topology*. ACM transactions on graphics (TOG) 32 (2012) 1-22.

[58] D. Bergondo Cañás. *Evaluación del potencial de aceleración de un código profesional de simulación de estructuras de solidificación en aleaciones metálicas por medio de paralelización en CPU y/o GPU*. Master Thesis, Universidade da Coruña (2024).

[59] D. Tourret, and A. Karma. *Growth competition of columnar dendritic grains: A phase-field study*. Acta Materialia 82 (2015) 64-83.

[60] D. Tourret, Y. Song, A.J. Clarke, and A. Karma. *Grain growth competition during thin-sample directional solidification of dendritic microstructures: A phase-field study*. Acta Materialia 122 (2017) 220-235.

[61] H. Esaka. *Dendrite growth and spacing in succinonitrile-acetone alloys*. PhD thesis, Ecole Polytechnique Fédérale de Lausanne, Switzerland (1986).

[62] E. Dorari, K. Ji, G. Guillemot, Ch.-A. Gandin, and A. Karma. *Growth competition between columnar dendritic grains–The role of microstructural length scales*. Acta Materialia 223 (2022) 117395.

[63] T. Takaki, S. Sakane, M. Ohno, Y. Shibuta, T. Aoki, and Ch.-A. Gandin. *Competitive grain growth during directional solidification of a polycrystalline binary alloy: Three-dimensional large-scale phase-field study*. Materialia, 1 (2018) 104-113.

[64] S. McFadden, D.J. Browne. *A generalized version of an Ivantsov-based dendrite growth model incorporating a facility for solute measurement ahead of the tip*. Computational Materials Science 55 (2012) 245-254.

[65] M. Rebow and D.J. Browne. *On the dendritic tip stability parameter for aluminium alloy solidification*. Scripta Materialia 56 (2007) 481-484.

[66] S. Akamatsu, and T. Ihle. *Similarity law for the tilt angle of dendrites in directional solidification of non-axially-oriented crystals*. Physical Review E 56 (1997) 4479.

[67] J. Deschamps, M. Georgelin, and A. Pocheau. *Growth directions of microstructures in directional solidification of crystalline materials*. Physical Review E 78 (2008) 011605.

[68] Y. Song, F.L. Mota, D. Tourret, K. Ji, B. Billia, R. Trivedi, N. Bergeon, and A. Karma. *Cell invasion during competitive growth of polycrystalline solidification patterns*. Nature Communications 14 (2023) 2244.

[69] T. Takaki, M. Ohno, Y. Shibuta, S. Sakane, T. Shimokawabe, and T. Aoki. *Two-dimensional phase-field study of competitive grain growth during directional solidification of polycrystalline binary alloy*. Journal of Crystal Growth 442 (2016) 14-24.

[70] Ch.-A. Gandin. *Experimental Study of the Transition from Constrained to Unconstrained Growth during Directional Solidification*. ISIJ International 40 (2012) 971-979.

[71] A. Otto, R. Gómez Vázquez, U. Hartel and S. Mosbah. *Numerical analysis of process dynamics in laser welding of Al and Cu*. Procedia CIRP 74 (2018) 691-695.

[72] A. Otto, R. Gómez Vázquez. *Fluid dynamical simulation of high speed micro welding*. J. Laser Appl. 30 (2018) 032411.

[73] R. Gómez. Vázquez, H. Koch, A. Otto. *Multi-physical simulation of laser welding*. Physics Procedia 56 (2014) 1334–1342.

[74] M. Buttazzoni, C. Zenz, A. Otto, R. Gómez Vázquez, G. Liedl and J. L. Arias. *A Numerical Investigation of the Laser Beam Welding of Stainless Steel Sheets with a Gap*. Applied Sciences, 11 (2021) 2549.

[75] C. Zenz, M. Buttazzoni, M. Martínez Ceniceros, R. Gómez Vázquez, J.R. Blasco Puchades, L. Portolés Griñán, and A. Otto. *Simulation-based process optimization of laser-based powder bed fusion by means of beam shaping*. Additive Manufacturing 77 (2023) 103793.

[76] E. Beyer. *Schweißen mit Laser*. Springer, Berlin, Heidelberg, New York (1995).

[77] H.L. Wie, J.W. Elmer, T. DebRoy. *Crystal growth during keyhole mode laser welding*. Acta Materialia 133 (2017) 10-20.

[78] A. Artinov, V. Karkhin, X. Meng, M. Bachmann, M. Rethmeier. A General *Analytical Solution for Two-Dimensional Columnar Crystal Growth during Laser Beam Welding of Thin Steel Sheets*. Applied Sciences 13 (2023) 6249.

[79] Y. Miyata, M. Okugawa, Y. Koizumi, T. Nakano. *Inverse columnar-equiaxed transition (CET) in 304 and 316L stainless steels melt by electron beam for additive manufacturing (AM)*. Crystals 11 (2021) 856.

[80] Z. Yang, A. Bauereiß, M. Markl, C. Körner. *Modeling laser beam absorption of metal alloys at high temperatures for selective laser melting*. Advanced Engineering Materials, 23 (2021) 2100137.

[81] K. Museth. NanoVDB: *A GPU-friendly and portable VDB data structure for real-time rendering and simulation*. In: ACM SIGGRAPH 2021 Talks (2021) 1-2.

# A particle-based approach for the prediction of grain microstructures in solidification processes

## SUPPLEMENTARY DATA

Salem Mosbah[1*], Rodrigo Gómez Vázquez[2,3], Constantin Zenz[2], Damien Tourret[4], Andreas Otto[2]

[1] *SOLIDIFICATION SAS, Sophia Antipolis, France*

[2] *E311-02 Research Unit of Photonic Technologies, Institute of Production Engineering and Photonic Technologies, TU Wien, Getreidemarkt 9/BA09, Vienna, Austria*

[3] *LKR Light Metals Technologies, Austrian Institute of Technology, Giefinggasse 2, Vienna, Austria*

[4] *IMDEA Materials, Madrid, Spain*

[*] *Corresponding author email: smosbah@solidification.io*

**Contents**

# 1. Unidirectional solidification of an Al-Si alloy

| Parameter | Symbol | Value | Unit |
|---|---|---|---|
| Domain size $\vec{e}_x$ direction | $L_x$ | 100 | mm |
| Domain size $\vec{e}_y$ direction | $L_y$ | 5 | mm |
| Domain size $\vec{e}_z$ direction | $L_z$ | 5 | mm |
| Finite Volume Cell size | $\Delta x = \Delta y = \Delta z$ | 1 | mm |
| Microstructure length scale (SDAS) | $\lambda_0$ | 50 | µm |
| Density | $\rho$ | 2600 | kg m$^{-3}$ |
| Specific heat capacity | $c_p$ | 1000 | J kg$^{-1}$ K$^{-1}$ |
| Thermal conductivity | $\kappa$ | 70 | J s$^{-1}$ m$^{-1}$ K$^{-1}$ |
| Dynamic viscosity | $\mu$ | 0.001 | kg m s$^{-1}$ |
| Latent heat of fusion | $\Delta H_f$ | 365 384.62 | J kg$^{-1}$ K$^{-1}$ |
| Melting temperature of pure Al | $T_m$ | 936.65 | K |
| Liquidus slope | $m_L$ | -6.5 | K$^{-1}$ |
| Initial concentration of Si | $w_0$ | 7.0 | wt.% |
| Eutectic concentration of Si | $w_{eut}$ | 13.31 | wt.% |
| Partition coefficient | $k$ | 0.13 | - |
| Diffusion coefficient in liquid | $D_l$ | 4.37e-09 | m$^2$ s$^{-1}$ |
| Diffusion coefficient in solid | $D_s$ | 10.0e-13 | m$^2$ s$^{-1}$ |
| Initial nucleus position | $(r_{i,x}, r_{i,y}, r_{i,z})$ | (0, 2.5, 2.5) | mm |
| Initial nucleus orientation | $(\psi_i, \Theta_i, \varphi_i)$ | (0, 0, 0) | ° |
| Lower boundary external temperature | $T_{ext}$ | 373.15 | K |
| Lower boundary heat transfer coefficient | $\alpha$ | 500 | W m$^{-2}$ K$^{-1}$ |
| Initial Temperature in domain | $T_0$ | 1073.15 | K |
| Kinetic model coefficient | $a$ | 2.9e-6 | m s$^{-1}$ K$^{-b}$ |
| Kinetic model exponent | $b$ | 2.7 | - |
| Time step | $\Delta t$ | 0.5 | s |

*Table S1: Domain size, discretization, thermophysical properties, initial conditions, boundary conditions, kinetic model for the 1D Al-7wt.%Si solidification case (Section 3.2 of the main text).*

## 2. Directional solidification of a quasi-2D thin sample

| Parameter | Symbol | Value | Unit |
|---|---|---|---|
| Domain size $\vec{e}_x$ direction | $L_x$ | 4 | mm |
| Domain size $\vec{e}_y$ direction | $L_y$ | 2 | mm |
| Domain size $\vec{e}_z$ direction | $L_z$ | 400 | µm |
| Finite Volume Cell size | $\Delta x = \Delta y = \Delta z$ | 400 | µm |
| Microstructure length scale (SDAS) | $\lambda_0$ | 192 … 12 | µm |
| Time step | $\Delta t$ | 0.1 | s |
| Liquidus Temperature | $T_l$ | 937.65 | K |
| Solidus Temperature | $T_s$ | 800 | K |
| Initial Temperature | $T_0$ | 938.65 | K |
| Thermal Gradient | $\nabla T$ | 1900 | K m$^{-1}$ |
| Pulling velocity | $v_{pull}$ | 86 | µm s$^{-1}$ |
| Cooling rate | $dT/dt$ | -0.1634 | K s$^{-1}$ |
| Lower boundary Temperature | $T_{lower}$ | $937.65 + t \cdot dT/dt$ | K |
| Upper boundary Temperature | $T_{upper}$ | $952.85 + t \cdot dT/dt$ | K |
| Kinetic model coefficient 1 | $a_1$ | 8.26e-6 | m s$^{-1}$ K$^{-2}$ |
| Kinetic model exponent 1 | $b_1$ | 2 | - |
| Kinetic model coefficient 2 | $a_2$ | 8.18e-5 | m s$^{-1}$ K$^{-3}$ |
| Kinetic model exponent 2 | $b_2$ | 3 | - |

*Table S2: Domain size, discretization, thermophysical properties, initial conditions, boundary conditions, kinetic model for the 2D grain growth competition case (Section 3.3 of the main text).*

## 3. Directional solidification of three-dimensional polycrystal

| Parameter | Symbol | Value | Unit |
|---|---|---|---|
| Domain size $\vec{e}_x$ direction | $L_x$ | 768 | µm |
| Domain size $\vec{e}_y$ direction | $L_y$ | 768 | µm |
| Domain size $\vec{e}_z$ direction | $L_z$ | 13.824 | mm |
| Finite volume cell size | $\Delta x = \Delta y = \Delta z$ | 16 | µm |
| Microstructure length scale | $\lambda_0$ | 24, 10 | µm |
| Macro time step | $\Delta t$ | 0.001 | s |
| Pulling velocity | $v_p$ | 100 | µm s$^{-1}$ |
| Thermal gradient | $G$ | 10, 100 | K mm$^{-1}$ |
| Melting temperature of pure Al | $T_m$ | 933.25 | K |
| Liquidus slope | $m$ | -2.668 | K wt.%$^{-1}$ |
| Initial concentration of Cu | $w_0$ | 3.01 | wt.%$^{-1}$ |
| Partition coefficient | $k$ | 0.14 | - |
| Gibbs-Thomson coefficient | $\Gamma$ | 0.24E-06 | K m |
| Interface energy anisotropy strength | $\varepsilon$ | 0.02 | - |
| Diffusion coefficient in liquid | $D_l$ | 3.0E-09 | m$^2$ s$^{-1}$ |
| Diffusion coefficient in solid | $D_s$ | 3.0E-13 | m$^2$ s$^{-1}$ |
| Kinetic model coefficient | $a$ | 1.834E-04 | m s$^{-1}$ K$^{-b}$ |
| Kinetic model exponent | $b$ | 2.282 | - |

*Table S3: Domain size, discretization, thermophysical properties, initial conditions, boundary conditions, kinetic model for the 3D grain growth competition case (Section 3.4 of the main text).*

## 4. Columnar-to-equiaxed transition in an Al-Si alloy

| Parameter | Symbol | Value | Unit |
|---|---|---|---|
| Domain height ($\vec{e}_z$ direction) | $L_z$ | 173 | mm |
| Domain diameter ($\vec{e}_x$-$\vec{e}_y$ direction) | $L_{xy}$ | 70 | mm |
| Number of Finite Volume cells $\vec{e}_z$ | $n_z$ | 100 | - |
| Number of Finite Volume cells $\vec{e}_x$- $\vec{e}_y$ | $n_x = n_y$ | 15 | - |
| Microstructure length scale | $\lambda_0$ | 250 | μm |
| Density of Al | $\rho_{Al}$ | **Table S5** | |
| Specific heat capacity | $c_p$ | **Table S6** | |
| Thermal conductivity | $\kappa$ | 170 | J s$^{-1}$ m$^{-1}$ K$^{-1}$ |
| Dynamic viscosity | $\mu$ | 0.001 | kg m s$^{-1}$ |
| Latent heat of fusion | $\Delta H_m$ | 400844 | J kg$^{-1}$ K$^{-1}$ |
| Melting temperature of pure Al | $T_m$ | 936.65 | K |
| Liquidus slope | $m_l$ | -6.5 | K$^{-1}$ |
| Initial concentration of Si | $w_0$ | 7.0 | wt.%$^{-1}$ |
| Eutectic concentration of Si | $w_{eutc}$ | 13.32 | wt.%$^{-1}$ |
| Density of Si | $\rho_{Si}$ | 2370 | kg m$^{-3}$ |
| Partition coefficient | $k$ | 0.13 | - |
| Diffusion coefficient in liquid | $D_l$ | 4.37e-09 | m$^2$ s$^{-1}$ |
| Diffusion coefficient in solid | $D_s$ | 10.0e-13 | m$^2$ s$^{-1}$ |
| Nuclei density lower boundary | $\rho_{Nuc}(z=0)$ | 5e05 | m$^{-2}$ |
| Nuclei density volume | $\rho_{Nuc}(z>0)$ | 1e09 | m$^{-3}$ |
| Lower boundary external temperature | $T(z=0)$ | [Carozzani 2012] | |
| Upper boundary heat flux (for t ≤ 900 s) | $Q$ | 3 000 | W m$^{-2}$ |
| Initial Temperature in domain | $T_0$ | 1017.1 | K |
| Kinetic model coefficient | $a$ | 2.9e-6 | m s$^{-1}$ K$^{-b}$ |
| Kinetic model exponent | $b$ | 2.7 | - |
| Time step | $\Delta t$ | 0.5 | s |

*Table S4: Domain size, discretization, thermophysical properties, initial conditions, boundary conditions, kinetic model for the 3D simulation of columnar-to-equiaxed transition in a cylindrical Al-Si ingot (Section 3.5 of the main text).*

| **Temperature** $T$ (K) | 850 | 850.15 | 855.15 | 860.15 | 865.15 | 870.15 | 875.15 | 880.15 | 885.15 | 890 | 891.15 |
|---|---|---|---|---|---|---|---|---|---|---|---|
| **Density** $\rho$ (kg m$^{-3}$) | 2535 | 2456 | 2451 | 2444 | 2437 | 2428 | 2418 | 2406 | 2392 | 2375 | 2370 |

*Table S5: Tabulated temperature-dependent values for density.*

| Temperature $T$ (K) | 573.15 | 673.15 | 773.15 | 823.15 | 850.15 | 855.15 | 860.15 | 865.15 |
|---|---|---|---|---|---|---|---|---|
| Thermal conductivity $\kappa$ (J s$^{-1}$ m$^{-1}$ K$^{-1}$) | 170 | 165 | 155 | 145 | 102.04 | 99.38 | 96.34 | 92.85 |
| | 870.15 | 875.15 | 880.15 | 885.15 | 890 | 891.15 | 1073.15 | |
| | 88.79 | 84.01 | 78.32 | 71.43 | 63.22 | 61 | 66 | |

*Table S6: Tabulated temperature-dependent values for thermal conductivity.*

## 5. Melting test case

| Parameter | Symbol | Value | Unit |
|---|---|---|---|
| Domain size $\vec{e}_x$ direction | $L_x$ | 1 | mm |
| Domain size $\vec{e}_y$ direction | $L_y$ | 2 | mm |
| Domain size $\vec{e}_z$ direction | $L_z$ | 0.1 | mm |
| Finite Volume Cell size | $\Delta x = \Delta y = \Delta z$ | 25 | µm |
| Microstructure length scale | $\lambda_0$ | 10 | µm |
| Liquidus Temperature | $T_l$ | 1723.15 | K |
| Solidus Temperature | $T_s$ | 1673.15 | K |
| Density | $\rho$ | 7873 | kg m$^{-3}$ |
| Specific heat capacity | $c_p$ | 450 | J kg$^{-1}$ K$^{-1}$ |
| Thermal conductivity | $\kappa$ | 300 | J s$^{-1}$ m$^{-1}$ K$^{-1}$ |
| Latent heat of fusion | $\Delta H_m$ | 1 | J kg$^{-1}$ K$^{-1}$ |
| Initial Temperature in domain | $T_0$ | 1674 | K |
| Cooling rate | $\partial T/\partial t$ | 20000 | K s$^{-1}$ |
| Nuclei density | $\rho_{nuc}$ | 1e13 | m$^{-3}$ |
| Kinetic model coefficient 1 | $a_1$ | 8.315e-6 | m s$^{-1}$ K$^{-b1}$ |
| Kinetic model coefficient 2 | $a_2$ | 9.628e-7 | m s$^{-1}$ K$^{-b2}$ |
| Kinetic model exponent 1 | $b_1$ | 2.490 | - |
| Kinetic model exponent 2 | $b_2$ | 3.622 | - |

*Table S7: Domain size, discretization, thermophysical properties, initial conditions, boundary conditions, kinetic model for the quasi-2D remelting test case (Section 4.1 of the main text).*

## 6. Laser beam welding of 316L stainless steel

| Parameter | Symbol | Value | Unit |
|---|---|---|---|
| Domain width | $L_x$ | 6 | mm |
| Domain length | $L_y$ | 40 | mm |
| Domain height | $L_z$ | 4 | mm |
| Finite volume cell size | $\Delta x = \Delta y = \Delta z$ | 125 | μm |
| Microstructure length scale | $\lambda_0$ | 20 | μm |
| Steel sheet thickness | $t_{sheet}$ | 2 | mm |
| Laser power | $P$ | 2.3 | kW |
| Welding speed | $v_{welding}$ | 20 | mm s$^{-1}$ |
| Laser spot size | $d_{laser}$ | 420 | μm |
| Initial grain structure undercooling | $\Delta T_{init}$ | 100 | K |
| Initial nuclei density | $\rho_{nuc,init}$ | 5.0E12 | m$^{-3}$ |
| Kinetic model coefficient 1 | $a_1$ | 8.315E-06 | m K$^{-b1}$ s$^{-1}$ |
| Kinetic model exponent 1 | $b_1$ | 2.49 | - |
| Kinetic model coefficient 2 | $a_2$ | 9.628E-07 | m K$^{-b2}$ s$^{-1}$ |
| Kinetic model exponent 2 | $b_2$ | 3.622 | - |
| Liquidus temperature | $T_l$ | 1708 | K |
| Solidus temperature | $T_s$ | 1675 | K |
| Boiling temperature | $T_b$ | 3134 | K |
| Latent heat of fusion | $\Delta H_m$ | 247.1E03 | J kg$^{-1}$ K$^{-1}$ |
| Latent heat of vaporization | $\Delta H_v$ | 6.21E06 | J kg$^{-1}$ K$^{-1}$ |
| Density (solid) | $\rho_s$ | 8000 | kg m$^{-3}$ |
| Density (liquid) | $\rho_l$ | 6936 | kg m$^{-3}$ |

*Table S8: Simulations parameters (growth kinetics coefficients, material parameters, process parameters, domain size and discretization) for the 3D welding case (Section 4.2 of the main text).*

| Parameter | Symbol | Temperature | Value | Unit |
|---|---|---|---|---|
| Specific heat capacity (solid) | $c_{p,s}$ | T = 300 K | 483.5 | J kg$^{-1}$ K$^{-1}$ |
| | | T = 1708 K | 684.0 | J kg$^{-1}$ K$^{-1}$ |
| Specific heat capacity (liquid) | $c_{p,l}$ | - | 800.0 | J kg$^{-1}$ K$^{-1}$ |
| Thermal conductivity (solid) | $\kappa_s$ | T = 300 K | 13.58 | J s$^{-1}$ m$^{-1}$ K$^{-1}$ |
| | | T = 1708 K | 32.6 | J s$^{-1}$ m$^{-1}$ K$^{-1}$ |
| Thermal conductivity (liquid) | $\kappa_l$ | T = 1708 K | 27 | J s$^{-1}$ m$^{-1}$ K$^{-1}$ |
| | | T = 3134 K | 42 | J s$^{-1}$ m$^{-1}$ K$^{-1}$ |
| Surface energy (liquid) | $\gamma_l$ | T = 1708 K | 1.802 | J m$^{-2}$ |
| | $d\gamma_l/dT$ | - | 3.66E-04 | J m$^{-2}$ K$^{-1}$ |
| Kinematic viscosity (liquid), Arrhenius pre-exponential factor | $\nu_0$ | - | 4.48E-08 | m$^2$ s$^{-1}$ |
| Kinematic viscosity (liquid), Arrhenius activation energy | $E_a$ | - | 45.5E-03 | J mol$^{-1}$ |

*Table S9: Temperature-dependent material properties for stainless steel 316L (linearly interpolated between data points).*

# 7. Nomenclature

| Symbol | Description | Unit |
|---|---|---|
| **Mathematical Symbols** | | |
| $a, b, a_1, b_1, a_2, b_2$ | Fitted Coefficients of Kinetic Law | - |
| $b_{w_i}$ | Solute Expansion Coefficient (for species $i$) | wt%$^{-1}$ |
| $c_p$ | Specific Heat Capacity | J kg$^{-1}$ K$^{-1}$ |
| $D$ | Diffusion Coefficient | m$^2$ s$^{-1}$ |
| $g$ | Volume Fraction | - |
| $\boldsymbol{g}$ | Gravity Acceleration (vectorial) | m s$^{-2}$ |
| $G$ | Thermal Gradient | K m$^{-1}$ |
| $H$ | Enthalpy | J kg$^{-1}$ |
| $\Delta H_f$ | Latent Heat of Freezing | J kg$^{-1}$ |
| $k$ | Partition Coefficient | - |
| $L$ | Length | m |
| $m_L$ | Liquidus Slope | K$^{-1}$ |
| $p$ | Pressure | kg m$^{-1}$ s$^{-2}$ |
| $r$ | Tip Radius | m |
| $t$ | Time | s |
| $\Delta t$ | Numerical Time Increment | s |
| $T$ | Temperature | K |
| $T_{eut}$ | Eutectic Temperature | K |
| $T_L$ | Liquidus Temperature | K |
| $T_{ref}$ | Reference Temperature | K |
| $\Delta T$ | Undercooling | K |
| $\dot{T}$ | Cooling Rate | K s$^{-1}$ |
| $\boldsymbol{U}$ | Fluid Velocity (vectorial) | m s$^{-1}$ |
| $v$ | Tip Growth Velocity | m s$^{-1}$ |
| $v_{pull}$ | Pulling Velocity | m s$^{-1}$ |
| $w$ | Concentration | - |
| $w_{l^*}$ | Concentration at Tip | - |
| $w_0$ | Nominal Concentration | - |
| $\varGamma$ | Gibbs-Thomson Coefficient | K m |
| $\kappa$ | Thermal Conductivity | W m$^{-1}$ K$^{-1}$ |
| $\lambda_0$ | Microstructural Length Scale | m |
| $\lambda_2$ | Secondary Dendrite Arm Spacing | m |
| $\mu$ | Dynamic Viscosity | kg m$^{-1}$ s$^{-1}$ |
| $\rho$ | Density | kg m$^{-3}$ |
| $\sigma^*$ | Tip Selection Parameter | - |
| $\psi, \theta, \varphi$ | Euler Angles | ° |

| | Superscripts | |
|---|---|---|
| *l* | Liquid | |
| *m* | Mushy | |
| *s* | Solid | |
| *si* | Internal Solid in Mushy Zone | |
| *t* | Value of Current Time Step | |
| $t_0$ | Value of Previous Time Step | |
| ∗ | Value of Previous Iteration | |
| | **Subscripts** | |
| *i, j* | Species of Multicomponent System | |
| *min* | Minimum | |
| *max* | Maximum | |
| *x, y, z* | Directions in Cartesian Coordinates | |
| *μ* | Mean Value | |
| *σ* | Standard Deviation | |
| | **Acronyms & Abbreviations** | |
| AM | Additive Manufacturing | |
| bcc | body centered cubic | |
| BC | Boundary Condition | |
| CA | Cellular Automaton | |
| CET | Columnar-to-Equiaxed Transition | |
| CFD | Computational Fluid Dynamics | |
| fcc | face centered cubic | |
| FE | Finite Element | |
| FV | Finite Volume | |
| GB | Grain Boundary | |
| hcp | hexagonal closest packed | |
| KGT | Kurz–Giovanola–Trivedi | |
| LBW | Laser Beam Welding | |
| NS | Navier-Stokes | |
| PDAS | Primary Dendrite Arm Spacing | |
| PF | Phase-Field | |
| SDAS | Secondary Dendrite Arm Spacing | |

*Table S10: List of symbols, superscripts, subscripts, and abbreviations.*